\documentclass[twocolumn]{aastex631}

\usepackage{multirow}
\usepackage{fancyvrb}
\received{January 31, 2023}
\revised{May 18, 2023}
\accepted{May 18, 2023}

\shorttitle{ASPIRED toolkit}
\shortauthors{Lam et al.}
\graphicspath{{./}{figures/}}
\begin{document}

\title{Automated SpectroPhotometric Image REDuction (\textsc{ASPIRED})}

\author[0000-0002-9347-2298]{Marco C. Lam}
\affiliation{School of Physics and Astronomy, Tel Aviv University, Tel Aviv 69978, Israel}
\affiliation{Astrophysics Research Institute, Liverpool John Moores University, IC2, LSP, 146 Brownlow Hill, Liverpool, L3 5RF, UK}
\affiliation{Astronomical Observatory, University of Warsaw, Al. Ujazdowskie 4, 00-478, Warszawa, Poland}

\author[0000-0003-3434-1922]{Robert J. Smith}
\affiliation{Astrophysics Research Institute, Liverpool John Moores University, IC2, LSP, 146 Brownlow Hill, Liverpool, L3 5RF, UK}

\author[0000-0001-7090-4898]{Iair Arcavi}
\affiliation{School of Physics and Astronomy, Tel Aviv University, Tel Aviv 69978, Israel}
\affiliation{CIFAR Azrieli Global Scholars program, CIFAR, Toronto, ON, M5G 1M1, Canada}

\author[0000-0001-8397-5759]{Iain A. Steele}
\affiliation{Astrophysics Research Institute, Liverpool John Moores University, IC2, LSP, 146 Brownlow Hill, Liverpool, L3 5RF, UK}

\author[0000-0003-2780-7843]{Josh Veitch-Michaelis}
\affiliation{Astrophysics Research Institute, Liverpool John Moores University, IC2, LSP, 146 Brownlow Hill, Liverpool, L3 5RF, UK}
\affiliation{Department of Physics and Wisconsin IceCube Particle Astrophysics Center, University of Wisconsin, Madison, WI 53706, USA}
\affiliation{ETH Z\"urich, Systems Group, Stampfenbachstrasse 114, 8092 Z\"urich, Switzerland}

\author[0000-0002-9658-6151]{Lukasz Wyrzykowski}
\affiliation{Astronomical Observatory, University of Warsaw, Al. Ujazdowskie 4, 00-478, Warszawa, Poland}


%% encouraged to surround collaboration identifiers with s. The 

\begin{abstract}
%%%%%%%%%%%%%%%%%%%%%%%%%%%%%%%%%%%%%%%%%%%%%%%%%%%%%%%%%%%%%%%%%%%%%%%%%%%%%%%%
We provide a suite of public open-source spectral data-reduction software to
rapidly obtain scientific products from all forms of long-slit-like
spectroscopic observations. Automated SpectroPhotometric
REDuction~(\textsc{ASPIRED}) is a \textsc{Python}-based spectral data-reduction
toolkit. It is designed to be a general toolkit with high flexibility
for users to refine and optimize their data-reduction routines for the
individual characteristics of their instruments. The default configuration is suitable for low-resolution long-slit spectrometers and provides a quick-look quality output. However, for repeatable science-ready reduced spectral data, some moderate one-time effort is necessary to modify the configuration. Fine-tuning and additional (pre)processing may be required to extend the reduction to systems with more complex setups. It is important to emphasize that although only a few parameters need updating, ensuring their correctness and suitability for generalization to the instrument can take time due to factors such as instrument stability.
We compare some
example spectra reduced with \textsc{ASPIRED} to published data processed with
\textsc{iraf}-based and \textsc{STARLINK}-based pipelines, and find no loss in
the quality of the final product. The \textsc{Python}-based, \textsc{iraf}-free
\textsc{ASPIRED} can significantly ease the effort of an astronomer in
constructing their own data-reduction workflow, enabling simpler solutions to
data-reduction automation. This availability of near-real-time science-ready
data will allow adaptive observing strategies, particularly important in, but
not limited to, time-domain astronomy.
\end{abstract}



\section{Introduction}
%%%%%%%%%%%%%%%%%%%%%%%%%%%%%%%%%%%%%%%%%%%%%%%%%%%%%%%%%%%%%%%%%%%%%%%%%%%%%%%%
With major global investments in multiwavelength and multimessenger surveys,
time-domain astronomy is entering a golden age. In order to maximally exploit
discoveries from these facilities, rapid spectroscopic follow-up observations
of transient objects~(e.g.,\ supernovae, gravitational-wave optical counterparts,
etc.) are needed to provide crucial {\em astrophysical} interpretations. 

Part of
the former OPTICON\footnote{\url{https://www.astro-opticon.org}; now the
OPTICON-RadioNet Pilot \url{https://www.orp-h2020.eu}} project coordinates the
operation of a network of largely self-funded European robotic and conventional
telescopes, collating common science goals and providing the tools to deliver
science-ready photometric and spectroscopic data in near real time. The goal is
to facilitate automated or interactive decision-making, allowing ``on-the-fly''
modification of observing strategies and rapid triggering of other facilities.
As part of the network's activity, a software development work package was
commissioned under the working title of Automated SpectroPhotometric
REDuction~(\textsc{ASPIRED}), coordinated on
\textsc{Github}\footnote{\url{https://github.com/cylammarco/ASPIRED}}.

%%%%%%%%%%%%%%%%%%%%%%%%%%%%%%%%%%%%%%%%%%%%%%%%%%%%%%%%%%%%%%%%%%%%%%%%%%%%%%%%
The ``industrial standard'' of spectral and image reduction is undoubtedly
the \textsc{iraf} framework~\citep{1986SPIE..627..733T, 1993ASPC...52..173T}.
It has powered many reduction engines in the past and present. However,
unfortunately, the National Optical Astronomy Observatory 
discontinued the support of the software in 2013, and it is now
entirely relying on community support\footnote{\url{https://iraf-community.github.io}}.
The \textsc{STARLINK}
library\footnote{\url{https://STARLINK.eao.hawaii.edu/STARLINK}} \citep{2014ASPC..485..391C, 2022ASPC..532..559B}
includes substantial resources for data-reduction tasks in the \textsc{Figaro} package. It was first initiated
in 1980 under the STARLINK Project, which became defunct in 2005, but the
software survived, and its maintenance effort has since transferred to the
East Asian Observatory. It also comes with a \textsc{Python}
wrapper\footnote{\url{https://github.com/STARLINK/STARLINK-pywrapper}}
to enable low-level code access from high-level programs. These software tools
have made significant contributions to the entire astrophysics community. They
are very often overlooked when they are so deeply embedded into many systems,
and users do not interact directly with them in this era of big data
and high data rates. However, such tools are not always available.
Many smaller facilities and individual users are still required to commission
their own data-reduction routines. Existing powerful libraries and
tools also come with a few drawbacks; the most notable obstacles concern the
ease of installation and linkage of libraries, which users often find difficult.

%%%%%%%%%%%%%%%%%%%%%%%%%%%%%%%%%%%%%%%%%%%%%%%%%%%%%%%%%%%%%%%%%%%%%%%%%%%%%%%%
In this generation of user-side astronomy data handling and processing, as
well as the emphasis on computing courses for scientists, \textsc{Python} is
among the most popular languages due to its ease to use with a shallow learning
curve, readable syntax and simple way to ``glue'' different pieces of software
together. Its flexibility to serve as a scripting and an object-oriented
language makes it useful in many use cases: developing visual tools
with little overhead, prototyping, web-serving, and compiling if
desired. While this broad range of functionality and high-level usage make it
relatively inefficient. \textsc{Python} is an excellent choice of
language for building wrappers on top of highly efficient and well-established codes.
Some of the most used packages, \textsc{scipy}~\citep{2020SciPy-NMeth}
and \textsc{numpy}~\citep{2020NumPy-Array}, are written in \textsc{Fortran}
and \textsc{C}, respectively, to deliver high performance. Multithreading
and multiprocessing are also possible with built-in and other third-party
packages, such as \textsc{mpi4py}~\citep{DALCIN20111124}. 

%%%%%%%%%%%%%%%%%%%%%%%%%%%%%%%%%%%%%%%%%%%%%%%%%%%%%%%%%%%%%%%%%%%%%%%%%%%%%%%%
Various efforts are being made to develop software for the current and
next generation of spectral data reduction. For example,
\textsc{PypeIt}~\citep{pypeit:zenodo, 2020JOSS....5.2308P} is designed for
tailor-made reduction for a range of instruments, and
\textsc{PyReduce}~\citep{2021A&A...646A..32P} is designed for optimal echelle
spectral reduction (but only handles sky subtraction if a sky frame is provided).
In the \textsc{Astropy} ``universe''~\citep{astropy:2013, astropy:2018},
\textsc{specreduce}\footnote{\url{https://github.com/astropy/specreduce}}~\citep{pickering_timothy_2022_7007991} is likely to be the
next-generation, user-focused data-reduction package of choice, but it is still in early
stages of deployment at the time of writing.
\textsc{pyDIS}\footnote{\url{https://github.com/StellarCartography/pydis}} has
all the essential ingredients for reducing spectra but has been out of
maintenance since 2019, and \textsc{specutils}
handles spectral analysis and manipulation but not the reduction itself.

%%%%%%%%%%%%%%%%%%%%%%%%%%%%%%%%%%%%%%%%%%%%%%%%%%%%%%%%%%%%%%%%%%%%%%%%%%%%%%%%
In \textsc{ASPIRED}, we have a different vision of how and what should be
abstracted from the users. Instead of providing a ready-to-go ``black box''
suited for a specific instrument or observations made under particular conditions,
we provide a toolkit that is as general as possible for users to have a set of
high-level data-reduction building blocks with which process the data in the ways most
appropriate to their instruments and observations. This shifts more of the work
and maintenance to the user end, but allows rapid modification of the
data-reduction workflow to any changes in the instrumental
configuration (such as a detector being refitted, or the detector plane
shifted and rotated by a few pixels during engineering work) and to new instruments. \textsc{ASPIRED}
is thus especially useful for works that repeatedly make use of multiple
spectrographs requiring rapid follow-ups.
Using \textsc{ASPIRED} only requires basic coding skills acquired by most
graduate-level astronomers. Similar to \textsc{iraf},
\textsc{PypeIt} and other open-source software, we welcome and encourage users
to provide feedback as well as to contribute to the code base.

%%%%%%%%%%%%%%%%%%%%%%%%%%%%%%%%%%%%%%%%%%%%%%%%%%%%%%%%%%%%%%%%%%%%%%%%%%%%%%%%
We use the SPRAT spectrograph \citep{2014SPIE.9147E..8HP} mounted on the Liverpool Telescope~(LT) as our
``first-light'' instrument for development. \textsc{ASPIRED} is currently used
by the transient science research group at Tel Aviv University
for a proprietary spectral data-reduction pipeline of the Las Cumbres Observatory
FLOYDS instrument. It is also known to be used
for quick-look reduction of data obtained by the MISTRAL spectrograph
mounted on the 1.93\,m telescope at the Observatoire de
Haute-Provence\footnote{\url{http://www.obs-hp.fr/guide/mistral/MISTRAL_spectrograph_camera.shtml}}.
However, as mentioned, we aim to allow high-level tools for users to build and
fine-tune their pipelines to support a wide range of
configurations~\citep{2022ASPC..532..537L, marco_2021_4463569}. As of the time
of writing, we have used \textsc{ASPIRED} to reduce data from the
William Herschel Telescope Intermediate-dispersion Spectrograph and
Imaging System\footnote{\url{https://www.ing.iac.es/astronomy/instruments/isis}}~(WHT/ISIS)
and the Auxiliary-port CAMera~\citep[ACAM;][]{2008SPIE.7014E..6XB}, the Las Cumbres
Observatory FLOYDS~\citep[Las Cumbres/FLOYDS;][]{2013PASP..125.1031B}, the Gemini Observatory
Gemini Multi-Object Spectrographs's long-slit
mode~\citep[Gemini/GMOS-LS;][]{2004PASP..116..425H}, the Gran Telescopio Canarias
Optical System for Imaging and low-Intermediate-Resolution Integrated
Spectroscopy~\citep[GTC/OSIRIS;][]{2000SPIE.4008..623C}, the Telescopio
Nazionale Galileo Device Optimised for the LOw
RESolution\footnote{\url{http://www.tng.iac.es/instruments/lrs}}~\citep[TNG/DOLORES;][]{1999ldss.work..157M},
the Very Large Telescope FOcal Reducer/low dispersion Spectrograph 2~(VLT/FORS2); and the Southern Astrophysical Research Telescope Goodman High Throughput Spectrograph~\citep[SOAR/GHTS][]{2004SPIE.5492..331C}.
We show examples of these reductions in Section~\ref{sec:examples}\footnote{See
also \url{https://github.com/cylammarco/aspired-example}}, and excerpts of
example codes in Appendix~\ref{sec:appendix}.

%%%%%%%%%%%%%%%%%%%%%%%%%%%%%%%%%%%%%%%%%%%%%%%%%%%%%%%%%%%%%%%%%%%%%%%%%%%%%%%%
This article is organized as follows. Section \textsection\ref{sec:development}
covers the development and organization of the software. Then, we discuss the
details of the spectral image-reduction procedures in Sections~\textsection
\ref{sec:image_reduction} and \ref{sec:spectral_reduction}. In
Section~\textsection{\ref{sec:examples}} we show a few example spectra and
compare then with published reductions.
Section~\textsection{\ref{sec:distribution}}
explains how \textsc{ASPIRED} can be installed. Finally in 
Section~\textsection\ref{sec:summary}, we summarize and discuss future
plans.

%%%%%%%%%%%%%%%%%%%%%%%%%%%%%%%%%%%%%%%%%%%%%%%%%%%%%%%%%%%%%%%%%%%%%%%%%%%%%%%%
This article is not intended to serve as an API document\footnote{\url{https://aspired.readthedocs.io/en/latest}\label{rtd}}, nor as a review of the
various methods concerning spectral extraction. Only the high-level
descriptions and the scientific and mathematical technicalities that are
important to the data-reduction processes are discussed here.

%%%%%%%%%%%%%%%%%%%%%%%%%%%%%%%%%%%%%%%%%%%%%%%%%%%%%%%%%%%%%%%%%%%%%%%%%%%%%%%%
\section{Development and Structure of \textsc{ASPIRED}}
\label{sec:development}

One of the development goals of \textsc{ASPIRED} is to design a piece of
software that is as modular and portable as possible, such that
it is operable on Linux, Mac and Windows
systems (while software for astronomy and astrophysics are
strongly leaning toward Unix systems, Windows has the highest
market share worldwide; furthermore, demographically it is the common choice
among university-managed computing systems, students and lower-income countries).
On top of that, \textsc{ASPIRED} is designed to rely on as few external
dependencies as possible. The ones we do use are those that would require a
substantial programming effort to reproduce and have a proven track record of
reliability and/or plan to remain maintained in the foreseeable future. The
explicit top-level dependencies are --
\textsc{astros-crappy}~\citep{2001PASP..113.1420V, curtis_mccully_2018_1482019}
\textsc{Astropy}~\citep{astropy:2013, astropy:2018},
\textsc{ccdproc}~\citep{matt_craig_2017_1069648},
\textsc{numpy}~\citep{2020NumPy-Array}
\textsc{plotly}~\citep{plotly},
\textsc{rascal}~\citep{2020ASPC..527..627V},
\textsc{scipy}~\citep{2020SciPy-NMeth},
\textsc{spectresc}~\citep{2017arXiv170505165C, lam_marco_c_2023_7865549},
\textsc{specutils}~\citep{nicholas_earl_2023_7803739}, and
\textsc{statsmodels}~\citep{seabold2010statsmodels}.

We host our source code on \textsc{Github}, which provides version control
and other utilities to facilitate its development. It uses \textsc{git}\footnote{\url{https://git-scm.com}},
issue and bug tracking, high-level project management, and automation with \textsc{Github Actions} upon each \texttt{commit} for the following:

\begin{enumerate}
    \item Continuous integration to install the software on Linux,
    \textsc{Mac} and \textsc{Windows} systems, and then perform unit tests with
    \textsc{pytest}~\citep{pytest6.2} over most of the functions in multiple
    versions of \textsc{Python}.
    \item Generating test coverage reports with \textsc{coveralls}\footnote{\url{    https://coveralls.io/github/cylammarco/ASPIRED}} which identifies the lines in
    the source code that are missed from the tests to assist us in maximizing the test coverage.
    \item Continuous deployment through \textsc{PyPI}\footnote{\url{https://pypi.org/project/aspired}} that allows immediate availability of the
    latest numbered version.
    \item Version tracking of the depending packages with 
    pull request is generated automatically (and the software is
    tested upon any update of the dependencies).
    \item Generating API documentation powered by \textsc{sphinx}\footnote{\url{https://www.sphinx-doc.org/en/master}} which scans for the decorators
    and automatically formats the written documentation (docstrings) into
    interactive \textsc{html} pages that can also be exported as a PDF file, and are
    hosted on Read the Docs$^{\ref{rtd}}$.
\end{enumerate}

%%%%%%%%%%%%%%%%%%%%%%%%%%%%%%%%%%%%%%%%%%%%%%%%%%%%%%%%%%%%%%%%%%%%%%%%%%%%%%%%
The initial work package divided the project broadly into three high-level
independent components, which we now detail.

%%%%%%%%%%%%%%%%%%%%%%%%%%%%%%%%%%%%%%%%%%%%%%%%%%%%%%%%%%%%%%%%%%%%%%%%%%%%%%%%
\subsubsection*{Graphical User Interface\\(Not in active development)}
In the beginning, we commissioned a prototype of the graphical user
interface\footnote{\url{https://github.com/cylammarco/gASPIRED}} with
\textsc{Electron}\footnote{\url{https://www.electronjs.org}}, which wraps on
top of \textsc{ASPIRED} without needing to adapt the code. Such a setup works seamlessly
for a few reasons. First, \textsc{Python} is an interpreter; it can
execute in run time. It is hence straightforward to run a \textsc{Python} server to
interact with an \textsc{Electron} instance. Second,
\textsc{Plotly}\footnote{\url{https://plotly.com}} has a good balance in
generating interactive and static plots; it requires little effort to interact
with both \textsc{Python} and \textsc{Electron}. Third, there is a
\textsc{JavaScript} version of
\textsc{SAOImageDS9}\footnote{\url{https://sites.google.com/cfa.harvard.edu/saoimageds9}}, \textsc{JS9},
that allows rapid and easy interactive image manipulation with an interface
that astronomers are accustomed to~\citep{2003ASPC..295..489J, eric_mandel_2022_6675771}.
\textsc{Electron} and \textsc{JS9} work like a web service; this prototype also
shows that an \textsc{ASPIRED} graphical user
interface can be deployed as an online tool. Such a high level of
interactivity is, however, no longer in active development and is only served
as a one-off technology demonstration of its capability.

%%%%%%%%%%%%%%%%%%%%%%%%%%%%%%%%%%%%%%%%%%%%%%%%%%%%%%%%%%%%%%%%%%%%%%%%%%%%%%%%
\subsubsection*{Arc Fitting for the Pixel-Wavelength Relation (a Concurrent Development)}

The wavelength calibration process involves identifying distinctive emission
lines from an arc-lamp spectrum, to which a function such as a polynomial can be
fit to map the pixel position to the wavelength value. Manual calibration is
straightforward but tedious: first, a user has to locate the peaks of an arc
spectrum in pixel space. Then, they have to assign a wavelength to each peak
based on the emission lines present from the element(s) of the arc lamp.
Finally, the function to map pixel values to a set of wavelengths is found.
Typically a list of known strong calibration lines is used, as provided by the
instrument operators in the form of an atlas or a template calibration spectrum.
Automation via template matching is a viable and commonly used approach for
spectral calibration if the instrument configuration is
stable~\citep{2020JOSS....5.2308P}. However, this introduces a strong
dependency on instrument-specific properties (such as the
vacuum/contamination condition of the lamp, the combination of elements in the
lamp, the vignetting on the focal plane, the response as a function of wavelength
across the detector, and saturation issues). There is also noise in peak-finding
routines, for example due to detector noise or quantization~(e.g.\ not using
subpixel peak finding), as well as complications due to blended lines.
Providing a general, template-free tool for such a purpose for a range of
specifications is far from easy, but would be much more flexible for cases when
templates are unavailable or the instrument configuration changes often.

In \textsc{ASPIRED}, the wavelength calibration module is powered by the
RANdom SAmple Consensus~(RANSAC) Assisted Spectral CALibration ~(\textsc{rascal},
\citealt{2020zndo...4117517V, 2020ASPC..527..627V}, Veitch-Michaelis \& Lam, in preperation)
library, a concurrent software development project.
\textsc{rascal} does not require templates and is designed to
work with an arbitrary instrument configuration. Users are only
required to supply a list of calibration
lines~(wavelengths), a list of peaks~(pixels), and some knowledge~(e.g.\ minimum
and maximum wavelength limits) of the system to obtain the most appropriate
pixel-to-wavelength solution. \textsc{rascal} builds on the work by \citet{2018ApOpt..57.6876S}, which
searches only for plausible sets of arc line/peak correspondences
that follow a linear approximation. RANSAC~\citep{fischler_bolles_1981}
is used to fit a higher-order polynomial model to the candidate correspondences
robustly. \textsc{rascal} extends this strategy to consider the top $N$ sets
of linear relations simultaneously: for each peak~(pixel), the most common
best-fit atlas line~(wavelength) is chosen from those $N$ subsets that have
a merit function better than a given threshold. This acts like a piecewise
linear fit and allows us to extract most of the correct matches from both the
red and blue ends of the spectrum - a feature that was
not available in \citet{2018ApOpt..57.6876S}.

%%%%%%%%%%%%%%%%%%%%%%%%%%%%%%%%%%%%%%%%%%%%%%%%%%%%%%%%%%%%%%%%%%%%%%%%%%%%%%%%
\subsubsection*{Data Reduction and Other Calibrations\\(This Work)}
This is the core of \textsc{ASPIRED}, which mainly handles the spectral
extraction and calibration processes. At the user level, the software is
organized into two main parts: \texttt{image\_reduction} and
\texttt{spectral\_reduction}. They are designed to be as modular as possible.
Users are free to use third-party reduction tools to perform part(s) of 
the reduction process and return to continue the reduction with
\texttt{image\_reduction} and \texttt{spectral\_reduction}.

In addition, the data structure class \texttt{SpectrumOneD} handles the
data and metadata of the extracted information. This provides a uniform format
to store the metadata and the extracted data throughout the entire spectral
extraction process. Since multiple spectra can be extracted from a
\texttt{TwoDSpec} object~(e.g.,\ extended target or multiple targets aligned on
the slit), each extracted spectrum is stored in an independent
\texttt{SpectrumOneD} object that gets passed from a \texttt{TwoDSpec} object.
In astronomy, the handling of observational
data is almost exclusively done with FITS files, but they are not
human-readable. In view of that, \textsc{ASPIRED} provides input
and output handling of FITS files, and outputs in a CSV format. All the
higher-level functions are wrappers on top of a \texttt{spectrumOneD} object.

%%%%%%%%%%%%%%%%%%%%%%%%%%%%%%%%%%%%%%%%%%%%%%%%%%%%%%%%%%%%%%%%%%%%%%%%%%%%%%%%
\section{Image Reduction}
\label{sec:image_reduction}

\textsc{ASPIRED} only intends to provide image-reduction methods
for bias, dark, and flat corrections. It is sufficient for simple instrumental
setups, which are the primary targets of \textsc{ASPIRED},
e.g., a long-slit spectrograph with a rectilinear
two-dimensional spectrum illuminated on the detector plane, i.e., where the dispersion
and spatial directions are perfectly perpendicular to each other across the
entire detector plane.
In systems with more complex optical designs, for example
instruments with multiple nonparallel spectra illuminated on the
detector plane, \textsc{ASPIRED} is not immediately usable. Extra image
manipulation is needed to further transform the data into simpler forms that
\textsc{ASPIRED} can handle. In the FLOYDS reduction example below, we
demonstrate that it is possible to extract two nonparallel curved spectra
from a single two-dimensional image. This is possible only because the two
spectra  (first-order red and second-order blue) that are simultaneously
imaged by the detector can be cleanly separated before spectral reduction,
and each spectrum is treated independently.

This module allows mean-, median-, and sum-combining, subtracting and dividing
by dark, bias, and flat frames. The FITS header of the first
light frame provided is retained in the reduced product, while the headers
from the other frames are not stored -- only the file paths are appended to
the end of the FITS header. The FITS file is defaulted to contain an empty
\texttt{PrimaryHDU} with the process data stored in the first
\texttt{ImageHDU} extension as recommended in the FITS Standard 
document\footnote{\url{https://fits.gsfc.nasa.gov/fits_documentation.html}},
but it can also be stored directly in the \texttt{PrimaryHDU} as most users
opt for. Many forms of interactive and static images can be exported,
powered by the \textsc{Plotly} image renderer.

This module only allows basic image reduction; users have to rely on third-party
software to correct for any other effects before continuing the
reduction with \texttt{spectral\_reduction}.

%%%%%%%%%%%%%%%%%%%%%%%%%%%%%%%%%%%%%%%%%%%%%%%%%%%%%%%%%%%%%%%%%%%%%%%%%%%%%%%%
\section{Spectral Reduction}
\label{sec:spectral_reduction}

In the highest-level terms, \textsc{ASPIRED}'s spectral-reduction procedures
after image reduction include (i) identifying and tracing the spectrum(a),
(ii) correcting for image distortion, (iii) extracting the spectrum(a),
(iv) performing wavelength calibration of the science target and a standard
star, (v) computing the sensitivity function using the standard star,
(vi) performing atmospheric extinction correction; and, finally (vii) removing
the telluric absorption features, (viii) resampling, and (ix) exporting the data.
We omit a few other processes, which we believe should be considered as part of
the postextraction processes or spectral analysis and manipulation. There are
many sophisticated software packages with good and long track records to
handle these processes post hoc. For example, sky-glow subtraction with
\textsc{Skycorr}~\citep{2014A&A...567A..25N}, or telluric
absorption removal without supplementary calibration data with
\textsc{MoleFit}~\citep{2015A&A...576A..77S, 2015A&A...576A..78K}.

The \texttt{spectral\_reduction} is a large module that covers both the
two-dimensional and one-dimensional operations through the \texttt{TwoDSpec} and
\texttt{OneDSpec} classes, respectively. The former provides image
rectification, spectral tracing, spectral extraction of the target and the
arc, and locating the peaks of the arc lines. The latter performs
wavelength calibration (with the \texttt{WavelengthCalibration} class),
flux calibration (with the \texttt{FluxCalibration} class), atmospheric
extinction correction, telluric absorption correction, and
spectral resampling. Diagnostic images can be exported at each step of the
two- and one-dimensional operations. This also includes the wavelength
calibration diagnostic that wraps around \textsc{rascal} and uses its native
plotting capability. Since the flux calibration has to be performed
with a standard star spectrum, it is necessary for the \texttt{OneDSpec}
object to have information on both the target and the standard one-dimensional
spectra. Thus, a \texttt{OneDSpec} object keeps a dictionary of lists: --
\texttt{science\_spectrum\_list} and \texttt{standard\_spectrum\_list}. At the
time of writing, \textsc{ASPIRED} only supports wavelength calibration from one (total
stacked) arc frame and flux calibration from one (total stacked) standard star frame.

The following describes the technical details of the nine reduction steps mentioned above.
We use the following convention throughout this article: the spectral image is
dispersed on the detector plane such that the left side~(small pixel number)
corresponds to the blue part of the spectrum and the right side~(large pixel
number) to the red part in the dispersion direction, while the direction
perpendicular to it is the spatial direction.

\texttt{spectral\_reduction} accepts images that are rotated or flipped
compared to the description above, but extra arguments have to be provided
in the initialization step to set the data into the orientation as defined
above.

%%%%%%%%%%%%%%%%%%%%%%%%%%%%%%%%%%%%%%%%%%%%%%%%%%%%%%%%%%%%%%%%%%%%%%%%%%%%%%%%
\subsection{Spectral Identification and Tracing}
\label{sec:tracing}
\begin{figure}
    \centering
    \includegraphics[width=\columnwidth]{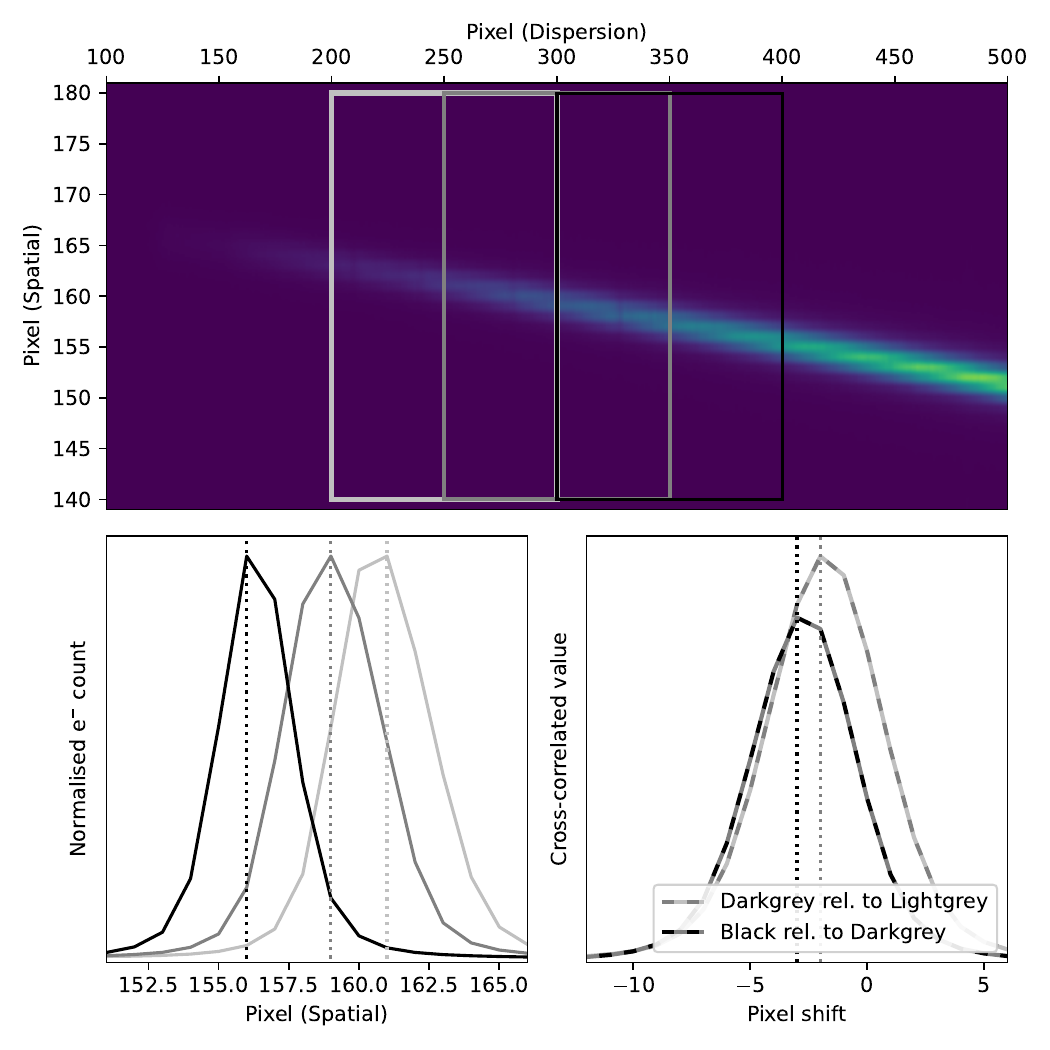}
    \caption{Top: a significantly tilted spectrum is used to illustrate how
    cross-correlation is used to find the shift between neighboring
    subspectra with a resampling factor of 1.0. The light gray, dark
    gray and black boxes are three example subspectra. Bottom left: the
    three subspectra are summed in the dispersion direction to generate
    the three profiles that are cross-correlated to compute the shift.
    Bottom right: the cross-correlated function for the light gray
    subspectrum relative to the dark gray subspectrum (plotted in
    alternating light and dark gray dashes), and the one for the black
    subspectrum relative to the dark gray subspectrum (plotted in
    alternating dark gray and black dashes).}
    \label{fig:trace}
\end{figure}

Conventionally, to obtain the position of a spectrum in a frame~(i.e.,\ a trace),
the center of the spectrum in the dispersion direction is first identified. Then,
a choice of algorithm scans across the spectral image to identify
the spectral position over the entire dispersion direction. In \textsc{ASPIRED},
the \texttt{ap\_trace} function works by dividing the two-dimensional spectrum
into subspectra. Then, each part is summed along the dispersion direction
before cross-correlating with the adjacent subspectra, using an optional variable
scaling factor that models a linear change in the width of the trace (this
optional scaling factor is turned off by default). This allows both the shift and the scaling of
the spectrum(a) along the dispersion direction to be found simultaneously. The
middle of the two-dimensional spectrum is used as the zero-point of the
procedure. This tracing procedure is illustrated in Figure \ref{fig:trace},
and described in more detail as follows:
\begin{enumerate}
    \item
        The input two-dimensional spectrum is divided into \texttt{nwindow}
        overlapping subspectra.
    \item
        Each subspectrum is summed along the dispersion direction
        in order to improve the signal(s) of the spectrum(a) along
        the spatial direction to generate the line-spread function
        of that subspectrum.
    \item
        Each line-spread function is upscaled by a factor of
        \texttt{resample\_factor} to allow for subpixel correlation
        precision. This utilizes the \texttt{scipy.signal.resample}
        function to fit the spatial profile with a cubic spline.
    \item
        The $i$th line-spread function is then cross-correlated to the $(i+1)$th
        line spread function. The shift and scale correspond to the maximum
        correlation within the tolerance limit, \texttt{tol}, and are stored.
    \item
        The shifts relative to the central subspectrum are used to
        fit for a polynomial solution with a constant arbitrary offset.
    \item
        While the spatial spectra are cross-correlated, they are also
        aligned and stacked to find the peak of the line-spread profile.
        Peak finding is performed with \texttt{scipy.signal.find\_peaks}
        and returns the list of peaks sorted by their prominence, which is
        defined as the maximum count of the peak subtracted by the local
        background. The total stacked line-spread profile is fitted with
        a Gaussian function to get the spread of the spectrum~(standard
        deviation). This would not be used in the case of top-hat
        extraction~(see section \textsection\ref{sec:tophat}).
\end{enumerate}

Manually supplying a trace (determined a priori or copied from another observation) is possible with the \texttt{add\_trace}
function. This is particularly useful for faint and/or confused
sources that are beyond the capability of automated tracing. For
example, if there is a faint transient source on top of a galaxy,
a manual offset may be needed for extraction if the host is
dominating the flux to a level that forced extraction is necessary.
For an optical design where the source spectrum is always illuminated
at the same pixels of a detector, it is possible to reuse the trace
from the standard frame for spectral extraction of the science target.

In the cross-correlation process, ASPIRED supports the upsampling of the data. 
Because more than one target could be incident on the slit, we do not attempt
to fit any profile at this stage. Hence, the shifts found are always
quantized to an integer pixel level. However, the polynomial fitted through
the shifts as a function of the dispersion direction is a continuous function.
The final precision depends on the upsampling rate. If no upsampling is
performed and the spectrum is tilted by one pixel across the entire range of
dispersion, for example, this step will not be able to get the tilt correctly.
By default, we set the upsampling rate to four. This should be sufficient to
correct for a tilt in the spectrum as little as 0.25 pixels across the entire
range of dispersion.

%%%%%%%%%%%%%%%%%%%%%%%%%%%%%%%%%%%%%%%%%%%%%%%%%%%%%%%%%%%%%%%%%%%%%%%%%%%%%%%%
\subsection{Image Rectification}
In some cases, a spectrum on the detector plane is not only tilted, but it is
also sufficiently distorted that the spatial and dispersion direction (in the
detector's Cartesian coordinates) are no longer orthogonal to each other. While
this does not cause any complication in the tracing process, the extraction has
to be performed along a curve so that the process is no longer a one-dimensional
process. Without proper weighting in extracting flux at the subpixel level,
significant under- or oversubtraction of the sky background can occur. More
complex methods have to be used to extract the spectra
directly~\citep[e.g.][]{2021A&A...646A..32P}, as implemented in
\textsc{PyReduce}, which can optimally extract highly distorted spectra.
However, \textsc{PyReduce} cannot perform sky subtraction using the ``wings''
in the sky region on either side of the line-spread profile. This has limited
the usage of \textsc{PyReduce} in typical single-frame extractions without
accompanying sky observations exposed using an identical slit setting. In
\textsc{ASPIRED}, a \texttt{TwoDSpec} object uses the spectral trace for
alignment in the spatial direction, and uses the sky emission lines, and
optionally coadded with the arc image, for alignment in the dispersion
direction. This process is to perfectly align the two-dimensional spectral coordinate system
with the detector's $x-y$ axes. Coadding the arc frame for the procedure can
improve the reliability of the rectification, as more bright lines are available
in the coadded image for computing the rectification function that shifts and
scales the image.

\begin{figure}
    \centering
    \includegraphics[width=\columnwidth]{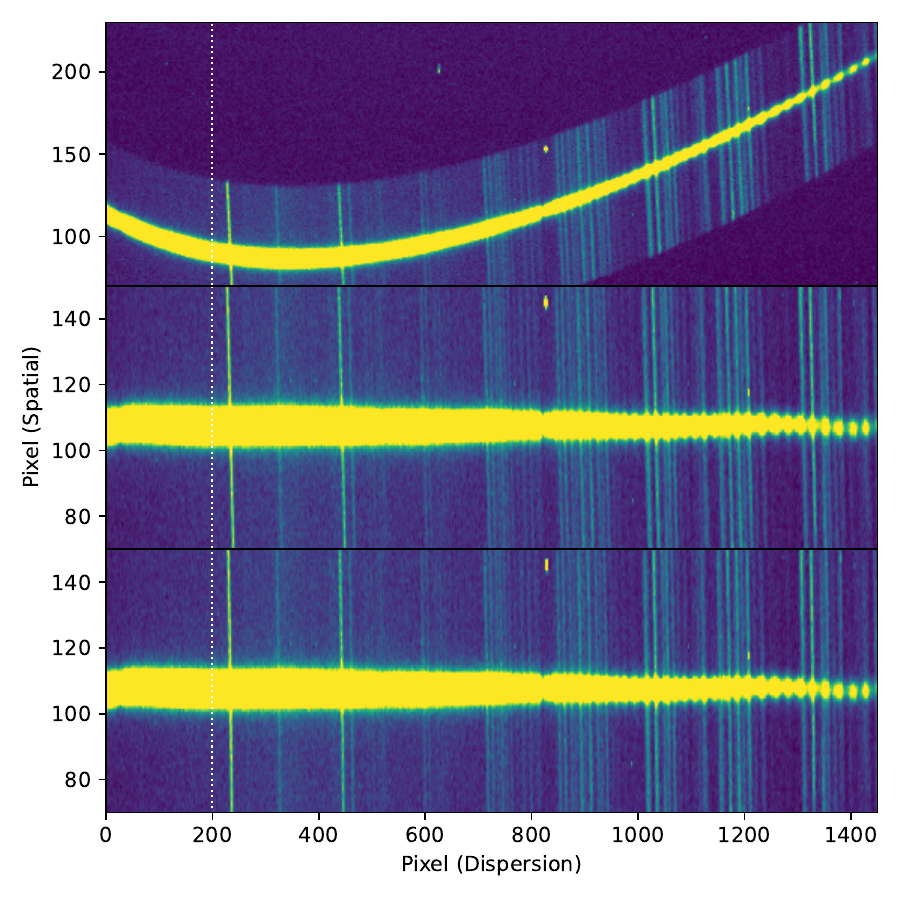}
    \caption{A two-dimensional spectrum from Las Cumbres/FLOYDS is used to demonstrate
    the rectification procedure. The dotted white line is aligned in the
    spatial direction to serve as a visual guide. Top: the trimmed image of a
    FLOYDS spectrum. Middle: the two-dimensional spectrum is resampled in the
    spatial direction based on the polynomial function of the trace. Bottom:
    the two-dimensional spectrum is resampled in the dispersion direction. The
    shifts and scales are found by cross-correlating the subspectra divided in
    the spatial direction. This is perpendicular to the tracing process.}
    \label{fig:rectify}
\end{figure}

The rectification is done independently first in the spatial direction and then
in the dispersion direction. In the spatial direction, the process only depends
on the trace. Each column of pixels is (scaled and) shifted by resampling it to
align to the center of the spectrum (middle panel of Fig.~\ref{fig:rectify}).
This process usually produces a spectrum that is tilted or curved in only the
dispersion direction. Therefore, a similar procedure has to be performed in
the dispersion direction. However, lines are not traced in this direction. In
order to find the size of the scaling and shifting, we repeat the tracing
procedures in Sec.~\ref{sec:tracing} from steps 1 through 5 in the
perpendicular~(dispersion) direction. A first- or second-order polynomial is
usually sufficient to fit the shift as a function of the number of pixels from
the trace. Then, the spectral image is resampled a second time but only in the
dispersion direction. The final result is shown in the bottom panel of
Fig.~\ref{fig:rectify}.

%%%%%%%%%%%%%%%%%%%%%%%%%%%%%%%%%%%%%%%%%%%%%%%%%%%%%%%%%%%%%%%%%%%%%%%%%%%%%%%%
At the time of writing, this process only works on a single trace. If
more than one trace is found/provided, only the one with the highest prominence
will be processed, where the 
prominence\footnote{\url{https://docs.scipy.org/doc/scipy/reference/generated/scipy.signal.peak_prominences.html}}
is defined as the maximum count of the peak subtracted by the local background. 
The resampling is performed with \textsc{spectresc}. It is possible to supply
a set of precomputed polynomials to perform the rectification, which can
significantly speed up the data-reduction process of a stable optical system.

%%%%%%%%%%%%%%%%%%%%%%%%%%%%%%%%%%%%%%%%%%%%%%%%%%%%%%%%%%%%%%%%%%%%%%%%%%%%%%%%
\subsection{Spectral Extraction}
\label{sec:extract}
\begin{figure}
    \centering
    \includegraphics[width=\columnwidth]{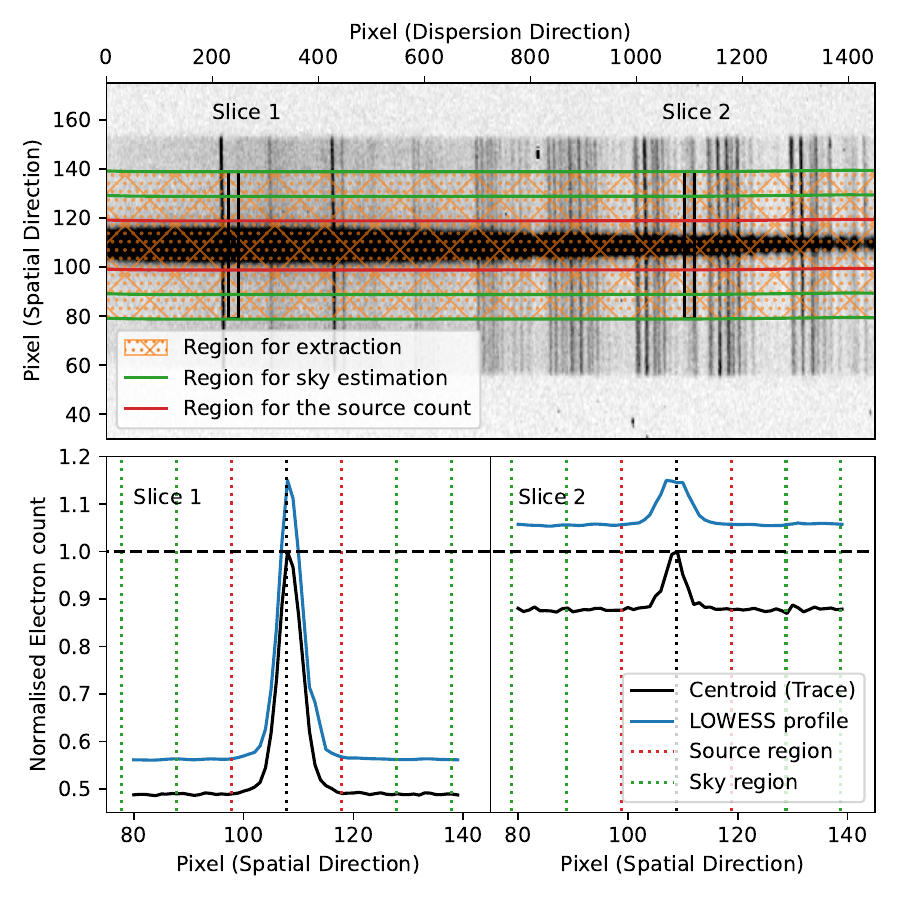}
    \caption{The same spectrum as in Figure~\ref{fig:rectify}. Top:
    the regions used for source and sky extractions are marked
    by the red and green boundaries. The light shade of orange
    hash marks the region that is used for spectral extraction.
    The two boxes show the two columns of pixels fitted with
    profiles in the bottom half of the figure (the boxes are inflated
    for clarity). Bottom: the independently normalized electron
    counts across the two slices; the much higher baseline on the right indicates a lower signal-to-background ratio. The vertical black dashed
    line is the centroid~(trace) of the spectrum, the two red dashed lines
    mark the regions of the source, and the pairs of green dashed lines on
    each side of the centroid show the regions used for sky extractions,
    respectively. The black line is the measured spectral profile (data), and
    the blue line is the fitted LOWESS line-spread function (model), offset for
    clarity. The right panel illustrates how an extraction over a nonrectified
    two-dimensional spectrum can lead to an increased sky-background level.}
    \label{fig:extraction}
\end{figure}

\begin{figure}
    \centering
    \includegraphics[width=\columnwidth]{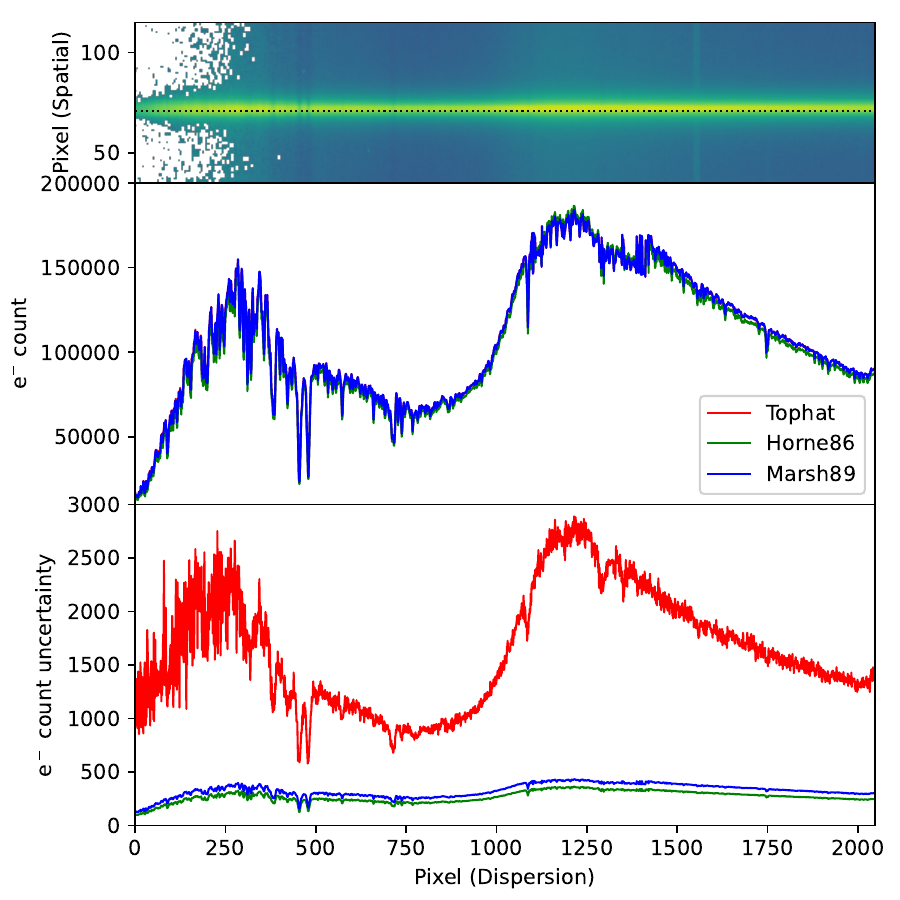}
    \caption{Top: the VLT/FORS2 spectrum and the trace of the standard
    star LTT\,7379. Middle: the extracted spectra in electron counts
    using the three extraction methods are very similar. Bottom: the
    uncertainties with the three extraction methods in units of electron counts.
    The top-hat method is clearly noisier than the two optimal methods.
    Extractions using the \texttt{Horne86} and \texttt{Marsh89} algorithms produce
    almost identical outputs.}
    \label{fig:extraction_compared}
\end{figure}

There are a few commonly used extraction methods; some work for all kinds of
spectral images, and some only work with specific observing strategies (e.g. the
flat-relative optimal extraction;~\citealt{2014A&A...561A..59Z}). The standard
textbook method is commonly called the top-hat extraction or the
normal extraction. It simply sums the electron counts over a given size
of the aperture and is robust and easy to use. However, this method does not
deliver the maximal signal-to-noise ratio~(S/N) from the available data. Various
optimal extraction algorithms can maximize the S/N. They work by down-weighting
the wings of the spectral profile, where almost all the photons come from the sky
background rather than the source~(see Fig.~\ref{fig:extraction}). The\
improvement is particularly significant for background-limited sources, which
are the case in most observations~(see Fig.~\ref{fig:extraction_compared}).
The extracted spectra and their associated uncertainties and sky-background
counts can be plotted for inspection. The residual image can also be exported
for diagnostics.

%%%%%%%%%%%%%%%%%%%%%%%%%%%%%%%%%%%%%%%%%%%%%%%%%%%%%%%%%%%%%%%%%%%%%%%%%%%%%%%%
\subsubsection*{Tophat/Normal Extraction}
\label{sec:tophat}
The top-hat extraction does not weight the pixels for extraction,
so every pixel has an equal contribution to the source count. Thus,
it is very robust in obtaining the total electron count across
a slice of pixels. The sky-background count can be extracted
from the regions outside the extraction aperture to be
subtracted from the spectrum.

%%%%%%%%%%%%%%%%%%%%%%%%%%%%%%%%%%%%%%%%%%%%%%%%%%%%%%%%%%%%%%%%%%%%%%%%%%%%%%%%
\subsubsection*{Horne-86 Optimal Extraction}
\citet[hereafter H86]{1986PASP...98..609H} is the golden standard
of optimal extraction of spectra from modern electronic detectors.
We follow the H86 process except for the profile modelling,
where we provide three options: the first is a fixed Gaussian
profile, the second uses LOcally Weighted Scatterplot
Smoothing~\citep[LOWESS;][]{doi:10.1080/01621459.1979.10481038}
regression to fit for a polynomial, and the third accepts
a manually supplied profile.

%%%%%%%%%%%%%%%%%%%%%%%%%%%%%%%%%%%%%%%%%%%%%%%%%%%%%%%%%%%%%%%%%%%%%%%%%%%%%%%%
The default option is to use the Gaussian fit because it is
more robust against noise, cosmic-ray contamination, and
artifacts. It is particularly useful in extracting a faint
spectrum when the image is dominated by background noise as the
Gaussian profile is constructed by fitting the line-spread
function from the total stack of all the
subspectra~(as described in Sec.~\ref{sec:tracing}).
LOWESS, on the other hand, would do a better job in fitting
the profile of a resolved galaxy. In any case, the quality
of the extraction from a valid profile should be at least as
good as that performed with the top-hat extraction method.

%%%%%%%%%%%%%%%%%%%%%%%%%%%%%%%%%%%%%%%%%%%%%%%%%%%%%%%%%%%%%%%%%%%%%%%%%%%%%%%%
\subsubsection*{Marsh-89 Optimal Extraction}
\citet[hereafter M89]{1989PASP..101.1032M} extends on the H86 algorithm by
fitting the change in the shape and centroid of the profile from one end of the
spectrum to the other. It is very suited for extracting a highly tilted
spectrum where the tilting direction is aligned with one direction in
the $x$-axis or the $y$-axis of the detector. The algorithm still relies on the
assumption that the spatial and dispersion directions are orthogonal across
the entire frame (in the detector pixel coordinates). In \textsc{ASPIRED}, we
adapt Prof. Ian Crossfield's set of public \textsc{python 2} code for
astronomy\footnote{\url{https://crossfield.ku.edu/python/}}
to \textsc{python 3}.

%%%%%%%%%%%%%%%%%%%%%%%%%%%%%%%%%%%%%%%%%%%%%%%%%%%%%%%%%%%%%%%%%%%%%%%%%%%%%%%%
\subsection{Wavelength Calibration}
As mentioned, the wavelength calibration is powered by \textsc{rascal}, which is
a concurrent development to this work. All the public functions from
\textsc{rascal} are available in \textsc{ASPIRED} so that users can have fine control
over the calibration. It works by applying a Hough transform to a set of
peak-line\,(pixel-wavelength) pairs (i.e.,~Hough pairs), and then identifying the
most probable set of Hough pairs that can be described by a polynomial
function which satisfies the required residual tolerance limit~(see more
details in \citealt{2020ASPC..527..627V})

\begin{figure}
    \centering
    \includegraphics[width=\columnwidth]{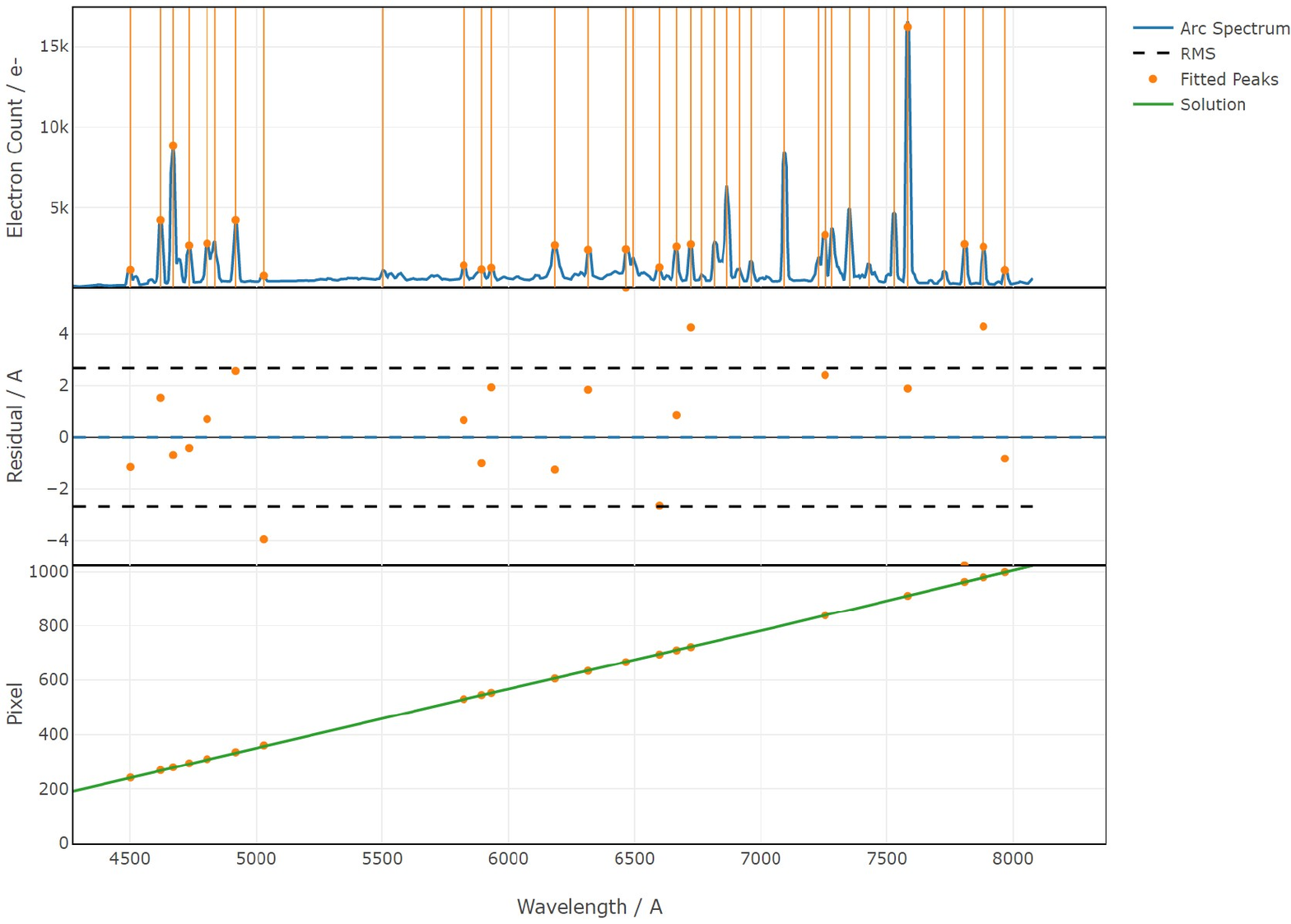}
    \caption{The native wavelength calibration diagnostic plots from \textsc{rascal}.
    Top: the arc is plotted in blue, the fitted peaks are marked by
    the orange dots. Middle: the residual plot shows the difference
    between the fitted peaks and the true wavelengths. Bottom: The
    pixel-wavelength function (green) is overplotted with the fitted
    peaks (orange).}
    \label{fig:wavecal}
\end{figure}

The diagnostic plots available are the set provided by \textsc{rascal}.
They include (i)~the spectrum of the arc lamp extracted using the traces from
the science and standard frames, wherein the peak of the arc lines are also marked;
(ii)~the Hough pairs and their constraints in the parameter space;
and (iii)~a plot showing the best-fit solution, the residual of the solution,
and the pixel-wavelength relation~(Fig.~\ref{fig:wavecal}).

Alternatively, the polynomial coefficients for the wavelength calibration can be
supplied manually, which would be useful for stable instruments in which the
variations in the dispersion are negligible. All the lines in the
range of 100 -- 30000 \AA\ 
available from the search form of the National Institute of Standards and
Technology Physical Measurement Laboratory are included, but we strongly
recommend providing arc lines manually since arc lamps, even of similar type,
vary between manufacturers and the conditions.

%%%%%%%%%%%%%%%%%%%%%%%%%%%%%%%%%%%%%%%%%%%%%%%%%%%%%%%%%%%%%%%%%%%%%%%%%%%%%%%%
\subsection{Flux Calibration}
\textsc{ASPIRED} is designed to address the common case where targets and
reference standards are observed contemporaneously in close succession. It is
well suited to constructing unsupervized, automated pipelines for stable and
repeatable instruments. In such cases, an instrument sensitivity function can be
created from several standard targets, stored and applied to many science target
observations. At the time of writing, \textsc{ASPIRED} cannot construct a
sensitivity function from multiple frames, but it can accept a user-supplied
polynomial function to apply the calibration.
 
The process of deriving a sensitivity function is by comparing a
wavelength-calibrated standard spectrum with the literature values to map the
electron count to flux as a function of wavelength. All the standard stars
available in \textsc{iraf}, on the Issac Newton Group of Telescopes web page,
and on the European Southern Observatory~(ESO) web page (Optical and UV
Spectrophotometric Standard Stars)
are available on \textsc{ASPIRED}\footnote{\url{https://cylammarco.github.io/SpectroscopicStandards/}}. We call the set of data a library,
and provide below the complete listing of sources and references for each of
the libraries where available.

\subsection*{European Southern Observatory}
The ESO standard spectra are grouped into five sets, they can be downloaded from
the ESO site.\footnote{\url{https://www.eso.org/sci/observing/tools/standards/spectra/}}.

\begin{enumerate}
    \item \texttt{esoctiostan} -- CTIO standards from \citet{1992PASP..104..533H, 1994PASP..106..566H}
    \item \texttt{esohststan} -- HST standards from \citet{1995AJ....110.1316B, 1996AJ....111.1743B}
    \item \texttt{esookestan} -- Oke standards from \citet{1990AJ.....99.1621O}
    \item \texttt{esowdstan} -- White dwarf standards from \citet{1995AJ....110.1316B}
    \item \texttt{esoxshooter} -- ESO VLT X-shooter standards from \citet{2014Msngr.158...16M, 2014A&A...568A...9M}
\end{enumerate}

\subsection*{Issac Newton Group of Telescopes}

The Issac Newton Group of Telescopes~(ING) listing is grouped into five sets by the last name of the authors\footnote{\url{http://www.ing.iac.es/Astronomy/observing/manuals/html_manuals/tech_notes/tn065-100/workflux.html}}.

\begin{enumerate}
    \item \texttt{ing\_oke} -- \citet{1990AJ.....99.1621O} standards
    \item \texttt{ing\_sto} -- \citet{1977ApJ...218..767S} standards
    \item \texttt{ing\_og} -- \citet{1983ApJ...266..713O} standards
    \item \texttt{ing\_mas} -- \citet{1988ApJ...328..315M} standards
    \item \texttt{ing\_fg} -- \citet{1984PASP...96..530F} standards
\end{enumerate}

\subsection*{\texttt{iraf} Standards}
The complete listing of \texttt{iraf} standards can be found online
\footnote{\url{https://github.com/iraf-community/iraf/tree/main/noao/lib/onedstds}}.
References are included where they are available.

\begin{enumerate}
    \item \texttt{irafblackbody}
    \item \texttt{irafbstdscal} -- KPNO IRS standards (i.e.\ those from the Henry Draper catalogue)
    \item \texttt{irafctiocal} -- The original CTIO standards \citet{1983MNRAS.204..347S, 1984MNRAS.206..241B}
    \item \texttt{irafctionewcal} -- The updated CTIO standards \citet{1992PASP..104..533H, 1994PASP..106..566H}
    \item \texttt{irafiidscal} -- KPNO IIDS standards \citet{1988ApJ...328..315M}
    \item \texttt{irafirscal} -- KPNO IRS standards \citet{1988ApJ...328..315M}
    \item \texttt{irafoke1990} -- HST standards \citet{1990AJ.....99.1621O}
    \item \texttt{irafredcal} -- KPNO IRS standards \& IIDS \citet{1988ApJ...328..315M} with wavelength beyond 8\,370\,$\mathrm{\AA}$
    \item \texttt{irafspechayescal} -- KPNO standards \citet{1988ApJ...328..315M}
    \item \texttt{irafspec16cal} -- CTIO standards \citet{1992PASP..104..533H, 1994PASP..106..566H} at 16$\mathrm{\AA}$ interval
    \item \texttt{irafspec50cal} -- KPNO standards \citet{1988ApJ...328..315M, 1990ApJ...358..344M} at 50$\mathrm{\AA}$ interval
\end{enumerate}

The calibration can be done in either AB magnitude or
flux density~(per unit wavelength). The two should give similar
results, but the response functions found would not be equivalent
because fitting to magnitudes is in logarithmic space and smoothing~(see
below) will have a different effect compared to flux fitting.

%%%%%%%%%%%%%%%%%%%%%%%%%%%%%%%%%%%%%%%%%%%%%%%%%%%%%%%%%%%%%%%%%%%%%%%%%%%%%%%%
\subsubsection*{Smoothing}
A Savitzky-Golay smoothing
filter\footnote{\url{https://docs.scipy.org/doc/scipy/reference/generated/scipy.signal.savgol_filter.html}}~\citep[hereafter, SG-filter]{1964AnaCh..36.1627S}
can be applied to the data before computing the sensitivity curve. This function
works by fitting low-order polynomials to localized subsets of the data to
suppress noise in the data at each point interval. It is similar to the
commonly used median boxcar filter but uses more weighted information to
retain information better while removing noise. This can be used independently
or with continuum fitting (see below). By default,
the smoothing is turned off, and when it is used it is defaulted to
remove only significant noise~(e.g.\ from unsubtracted cosmic-rays).

%%%%%%%%%%%%%%%%%%%%%%%%%%%%%%%%%%%%%%%%%%%%%%%%%%%%%%%%%%%%%%%%%%%%%%%%%%%%%%%%
\subsubsection*{Continuum Fitting}
It is optional for the sensitivity function to be derived from the continuum
of the standard spectrum. This removes absorption lines in the standard
spectrum. This process is strongly dependent on both the resolution of the
literature and the observed standards, as well as the absorption lines present
in them. The continuum is found using the \textsc{specutils}'s
\texttt{fits\_continuum}. This can remove any outlying random noise and
absorption lines when computing the  sensitivity. Users are reminded to be
cautious with this procedure as removing the absorption features in computing
the sensitivity function can significantly affect the flux calibration near
the absorption features.

%%%%%%%%%%%%%%%%%%%%%%%%%%%%%%%%%%%%%%%%%%%%%%%%%%%%%%%%%%%%%%%%%%%%%%%%%%%%%%%%
\subsubsection*{Sensitivity Function}
\begin{figure}
    \centering
    \includegraphics[width=\columnwidth]{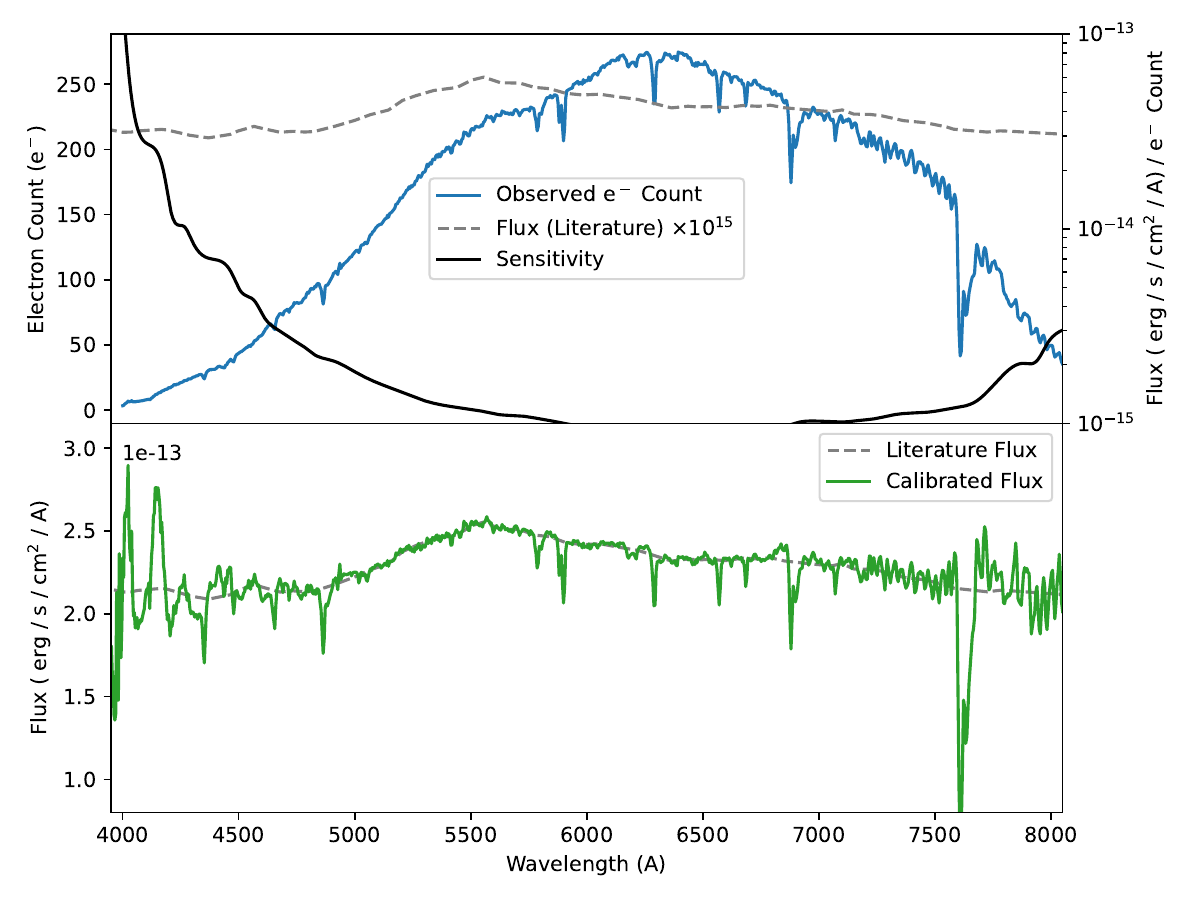}
    \caption{Top: the extracted standard spectrum in electron
    counts~(blue), the literature spectrum in units of flux density~(orange),
    and the sensitivity function~(y-axis on the right; black). Bottom:
    the literature spectrum (same as the one above) is plotted in orange, and
    the calibrated observed spectrum is plotted in green.}
    \label{fig:fluxcal}
\end{figure}

The sensitivity function is computed by dividing the observed standard spectrum
by the literature one and then interpolating the result using a spline or a
polynomial function~(Fig.~\ref{fig:fluxcal}). This can be done with or without
any presmoothing and/or continuum fitting. In case of a high-resolution
literature standard spectrum~(e.g.\ the ESO X-Shooter standards), the
absorption-line profiles should be used as part of the sensitivity computation,
while in the case of the standard stars from \citet{1990AJ.....99.1621O},
which removed absorption lines when producing the standard spectra, a
smooth continuum of the observed spectrum should be used for computing
the sensitivity function. Users can also manually supply a sensitivity function
as a callable function that accepts a wavelength value and returns the $\log$
of the sensitivity value at that wavelength. When such a function is provided
manually, it is assumed to be in the units of flux per second.

%%%%%%%%%%%%%%%%%%%%%%%%%%%%%%%%%%%%%%%%%%%%%%%%%%%%%%%%%%%%%%%%%%%%%%%%%%%%%%%%
\subsection{Atmospheric Extinction Correction}
\begin{figure}
    \centering
    \includegraphics[width=\columnwidth]{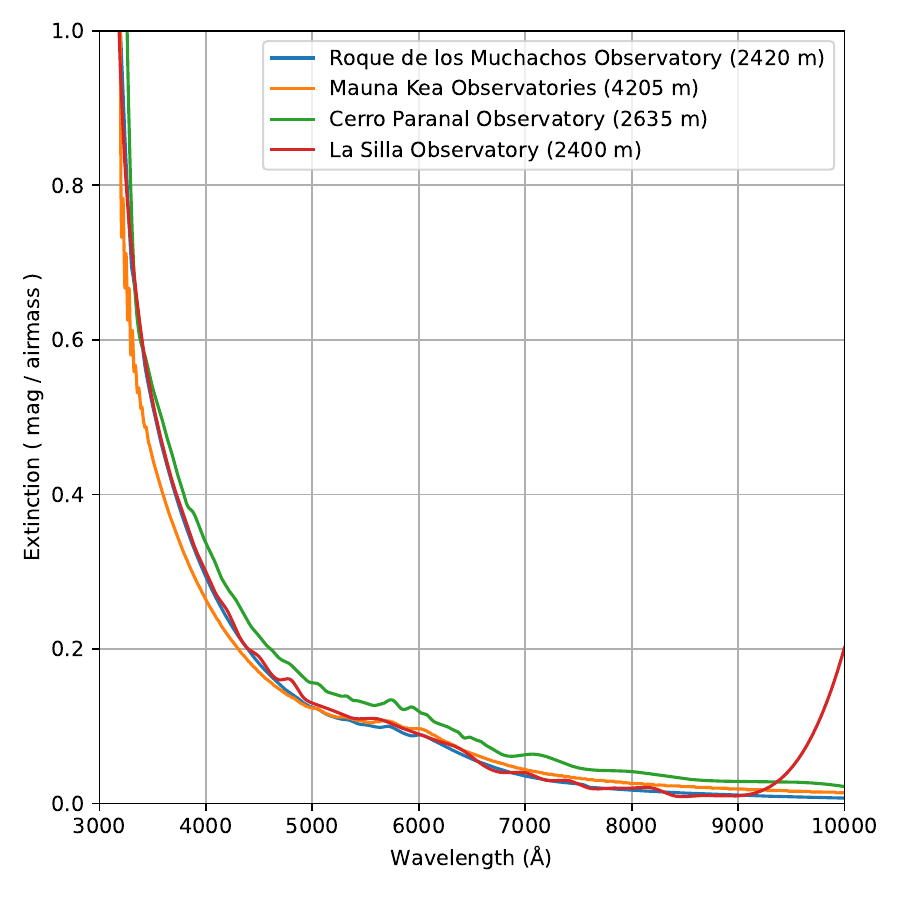}
    \caption{Extinction curves measured at Roque de los
    Muchachos Observatory~(2420\,m; blue), Mauna Kea Observatories~(4205\,m;
    orange), Cerro Paranal Observatory~(2635\,m; green), and La Silla
    Observatory~(2400\,m; red). The extinction table of La Silla Observatory
    terminates at $9000$\,\AA. The rise at the red end of the curve is an
    undesired artefact due to extrapolation using a cubic spline. We are
    explicitly plotting this range to serve as a warning since we opt to
    preserve the raw data as provided without appending fake data points.}
    \label{fig:extinction}
\end{figure}

\textsc{ASPIRED} currently has four built-in atmospheric extinction curves (Fig.~\ref{fig:extinction}):

\begin{enumerate}
    \item Roque de los Muchachos Observatory, Spain~(La Palma, 2420\,m)\footnote{\url{http://www.ing.iac.es/astronomy/observing/manuals/ps/tech\_notes/tn031.pdf}}.
    \item Mauna Kea Observatories, USA~\citep[Big Island, Hawaii, 4205\,m;][]{2013A&A...549A...8B}.
    \item Cerro Paranal Observatory, Chile~\citep[Atacama Desert, 2635\,m;][]{2011A&A...527A..91P}.
    \item La Silla Observatory, Chile~(Atacama Desert, 2400\,m)\footnote{\url{https://www.eso.org/public/archives/techdocs/pdf/report\_0003.pdf}}.
\end{enumerate}

These curves describe the extinction in magnitude per air mass as a function of wavelength,
roughly between $3000$ and $10000$\,\AA. Alternatively, a callable function
in the appropriate units can be supplied to perform the extinction correction.
The air mass of the observation can be found from the header if it is reported
using the conventional keywords for air mass.

%%%%%%%%%%%%%%%%%%%%%%%%%%%%%%%%%%%%%%%%%%%%%%%%%%%%%%%%%%%%%%%%%%%%%%%%%%%%%%%%
\subsection{Telluric Absorption Removal}
During the process of generating the sensitivity function, masks are used
over the telluric regions to generate the telluric absorption
profile from the standard star. The default masking regions are $6850-6960$
and $7580-7700$\ \AA\ only. This telluric profile can then be multiplied
by a factor to be determined in order to remove the telluric absorption
features in the science target. The best multiplicative factor is found
by minimizing the difference between the continuum and the
telluric-absorption-corrected spectrum simultaneously for both the sampled
telluric regions~(there are only two ranges in the default setting, but if more regions are
provided, this process will apply to all of them simultaneously).
An extra multiplier can be manually provided to adjust the
strength of the subtraction. This is designed for manually fine-tuning the
absorption factor; otherwise, it defaults to $1.0$.

%%%%%%%%%%%%%%%%%%%%%%%%%%%%%%%%%%%%%%%%%%%%%%%%%%%%%%%%%%%%%%%%%%%%%%%%%%%%%%%%
\subsection{Resampling}
All the operations above are performed as a function of the native detector
pixels. This includes the telluric absorption corrections as the telluric profile
is generated in the standard spectrum pixel/wavelength coordinates. The
correction applied to the target spectrum is performed using an interpolated
function of the profile at the wavelengths of the target spectrum. 

Many legacy software packages can only read the spectrum using the header information to create
the wavelength coordinates using only three parameters: the wavelength of the
first element of the spectral array, the bin width of each wavelength
coordinate, and the length of the spectral array. This requires a uniformly
sampled wavelength axis, which never happens naturally as there is always some
level of image distortion at the detector plane. A resampling is necessary
to turn the data into such a format.

The resampling is performed with the \textsc{spectresc} package. Although
this package allows nonuniform wavelength spacing resampling, we are only
using it for uniform resampling for the purpose of exporting the data as FITS
files.

%%%%%%%%%%%%%%%%%%%%%%%%%%%%%%%%%%%%%%%%%%%%%%%%%%%%%%%%%%%%%%%%%%%%%%%%%%%%%%%%
\subsection{Output}
At the time of writing, 22 types of output to CSV and FITS files are supported.
Each CSV has the various exported data (see below) stored in a separate columns, while the FITS has
the data stored in separate header data units (HDUs). Each spectrum is exported
as a separate file. At the time of writing, the header is not exported when the
output option is CSV. The options for output are as follows:

\begin{enumerate}
    \item \texttt{trace}: two columns/HDUs containing the pixel coordinates
    of the trace and its width in the spatial direction.
    \item \texttt{count}: three columns/HDUs containing the electron count
    of the spectrum, its uncertainty, and the sky background.
    \item \texttt{weight\_map}: one column/HDU for the extraction profile;
    in top-hat and \texttt{horne86} extraction this is a one-dimensional array,
    while the \texttt{marsh90} extraction returns a two-dimensional array.
    \item \texttt{arc\_spec}: three columns/HDUs for the one-dimensional arc
    spectrum, pixel coordinates of the arc line position (subpixel precision),
    and the pixel coordinates of the arc line effective position~(e.g.\ accounting for chip gaps) in the dispersion direction.
    \item \texttt{wavecal}: one column/HDU for the polynomial coefficients
    for wavelength calibration. The header contains the information regarding the
    polynomial type.
    \item \texttt{wavelength}: one column/HDU for the wavelength at the
    native pixel position.
    \item \texttt{sensitivity}: one column/HDU containing the sensitivity
    function of the detector as a function of the native detector pixel.
    \item \texttt{flux}: three columns/HDUs containing the
    flux of the spectrum, its uncertainty, and the sky background.
    \item \texttt{atm\_ext}: one column/HDU containing the atmospheric
    extinction correction factor at each wavelength position.
    \item \texttt{flux\_atm\_ext\_corrected}: three columns/HDUs containing
    the atmospheric-extinction-corrected flux of the spectrum, its
    uncertainty, and the sky background.
    \item \texttt{telluric\_profile}: one column/HDU containing the telluric
    absorption profile from the standard spectrum.
    \item \texttt{flux\_telluric\_corrected}: three columns/HDUs containing
    the telluric-absorption-corrected flux, uncertainty, and the sky
    background.
    \item \texttt{flux\_atm\_ext\_telluric\_corrected}: three columns/HDUs
    containing the atmospheric-extinction-corrected and telluric-absorption-corrected flux,
    uncertainty, and the sky background.
    \item \texttt{wavelength\_resampled}: one column/HDU containing the
    wavelength at each resampled position.
    \item \texttt{count\_resampled}: three columns/HDUs containing the
    electron count of the spectrum, its uncertainty, and the sky background
    being subtracted during the extraction process at the resampled wavelength.
    \item \texttt{sensitivity\_resampled}: one column/HDU containing the
    sensitivity function of the detector as a function of resampled wavelength.
    \item \texttt{flux\_resampled}: three columns/HDUs containing the
    flux of the spectrum, its uncertainty, and the sky background at the
    resampled wavelength.
    \item \texttt{atm\_ext\_resampled}: one column/HDU containing the
    atmospheric extinction correction factor at the resampled wavelength.
    \item \texttt{flux\_resampled\_atm\_ext\_corrected}: three columns/HDUs
    containing the atmospheric-extinction-corrected flux of the spectrum, its
    uncertainty, and the sky background at the resampled wavelength.
    \item \texttt{telluric\_profile\_resampled}: one column/HDU containing
    the telluric absorption profile from the standard spectrum at the resampled
    wavelength.
    \item \texttt{flux\_resampled\_telluric\_corrected}: three columns/HDUs
    containing the telluric-absorption-corrected flux, uncertainty, and the sky
    background at the resampled wavelength.
    \item \texttt{flux\_resampled\_atm\_ext\_telluric\_corrected}: three
    columns/HDUs containing the atmospheric-extinction-corrected and
    telluric-absorption-corrected flux, uncertainty, and the sky background
    at the resampled wavelength.
\end{enumerate}

%%%%%%%%%%%%%%%%%%%%%%%%%%%%%%%%%%%%%%%%%%%%%%%%%%%%%%%%%%%%%%%%%%%%%%%%%%%%%%%%
\section{Validation}
\label{sec:examples}
Being a flexible toolkit, \textsc{ASPIRED} is not designed to reduce data for a
specific instrument. Instead, it can be customized to suit many configurations,
as long as they are long-slit-like. We demonstrate this here with data products
from eight different instruments, most of which have conventional long-slit
settings. A summary of the following example reductions is tabulated in
Table~\ref{tab:summary}.

The purpose of the comparison made here is to demonstrate that
the reductions are sufficiently consistent with representative spectra in the literature
for a range of instruments and astrophysical sources. However, a full
systematic comparison is beyond the scope of this work. Such a level of
detailed validation might be required for any service providers who wish to
adopt \textsc{ASPIRED} as their official reduction tool, and will be
included in future work.

\subsection{Example Data Reduction Results}
In the case of Gemini/GMOS, a single spectrum is exposed onto
three detectors, and each detector is read out with four amplifiers. The bias 
subtraction is also unconventional, so all the image data-reduction procedures
have to be done outside of \textsc{ASPIRED} beforehand. We compare our reduction against
the official one for the kilonova AT 2017gfo\footnote{\url{https://www.wiserep.org/}}~\citep{2017ApJ...848L..32M}. 

Las Cumbres Observatory/FLOYDS has a beam splitter to expose the first-order
red beam and the second-order blue beam into two nonparallel
curved spectra on a single chip. Both the northern and southern FLOYDS suffer
from strong fringing effects. This is not handled natively through the API of 
\textsc{ASPIRED}. Instead, the correction was done by directly modifying the
electron \texttt{count} of the \texttt{OneDSpec} object. The fringe subtraction
workflow follows that of the official
pipeline\footnote{\url{https://github.com/LCOGT/floyds_pipeline}}. Here, the
Type II supernova iPTF14hls is used as an example for
comparison~\citep{2017Natur.551..210A}.

LT/SPRAT is a high-throughput low-resolution spectrograph with
a conventional optical design; in this case, we compare our reduction to that
published for a nova shell, DO Aql, from a stack of four individual 
exposures\footnote{\url{https://telescope.livjm.ac.uk/cgi-bin/lt_search}}~\citep{2020MNRAS.499.2959H}. 

The VLT/FORS2 data
of the helium accretor~(AM CVn) V418 Ser, stacked from 21 individual exposures, is
used as our last comparison~\citep{2020MNRAS.496.1243G}. 

\begin{figure*}
    \centering
    \includegraphics[width=\textwidth]{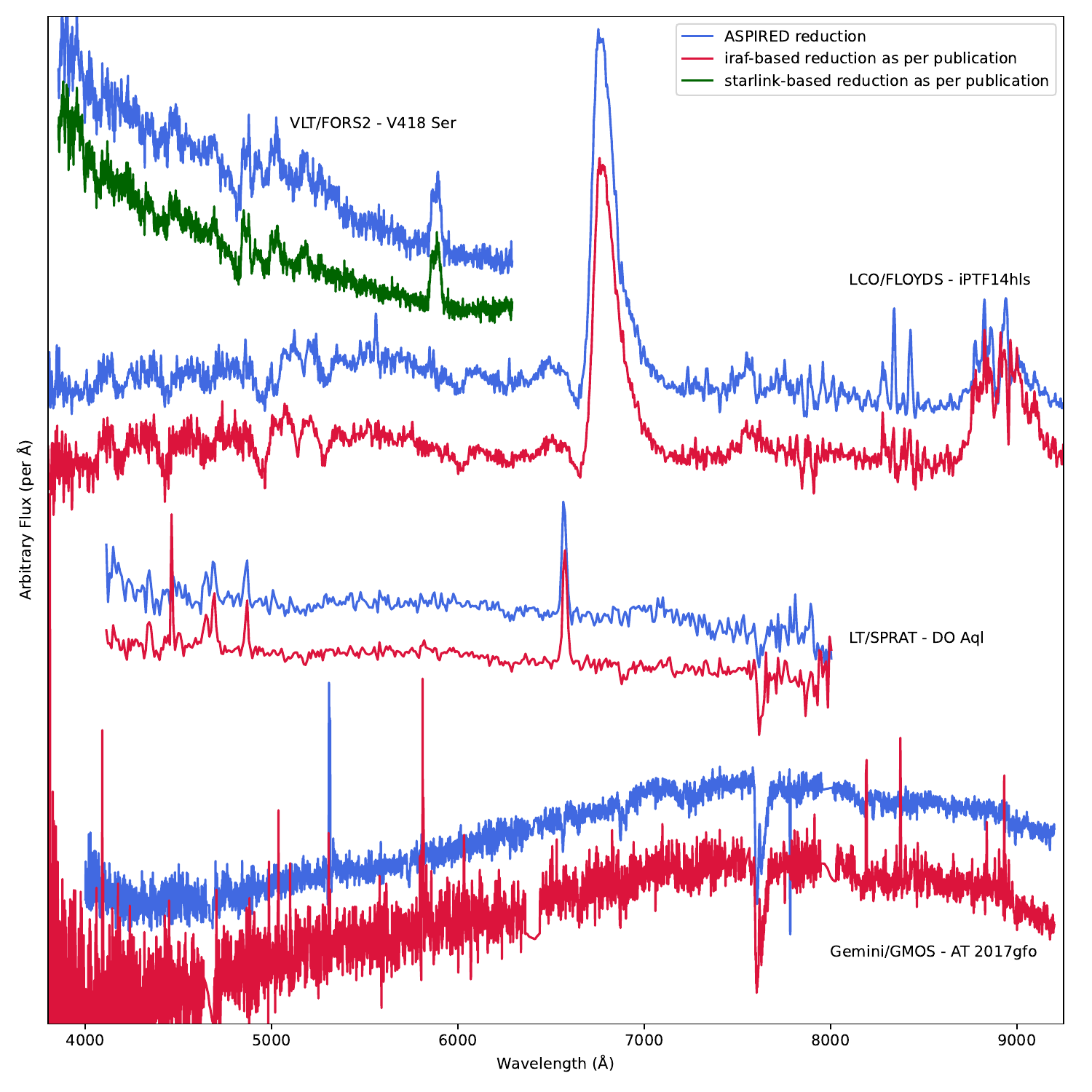}
    \caption{From top to bottom are the comparison spectra taken with VLT/FORS2,
    Las Cumbres/FLOYDS, LT/SPRAT and Gemini/GMOS in long-slit mode. The published FORS2
    spectrum was reduced using \textsc{Molly}~(green) powered by the \textsc{STARLINK}
    library. The other three (red) were reduced using the respective \textsc{iraf}-based
    pipelines. All spectra reduced with \textsc{ASPIRED}~(blue) are optimally extracted using the \cite{1986PASP...98..609H} algorithm and resampled to match
    the respective published spectra. Telluric correction is applied using the default
    setting.}
    \label{fig:reduction_compared}
\end{figure*}

The first three spectra were 
compared against data reduction from \textsc{iraf}-based pipelines, while the
last one is reduced with \textsc{Molly}\footnote{
\url{https://cygnus.astro.warwick.ac.uk/phsaap/software/molly/html/INDEX.html}}~\citep{2019ascl.soft07012M},
which is built on top of \textsc{STARLINK}. All of our reductions are resampled to match
the resolution in the equivalent published spectra. Our comparisons are shown in Figure \ref{fig:reduction_compared}.

\begin{figure}
    \centering
    \includegraphics[width=\columnwidth]{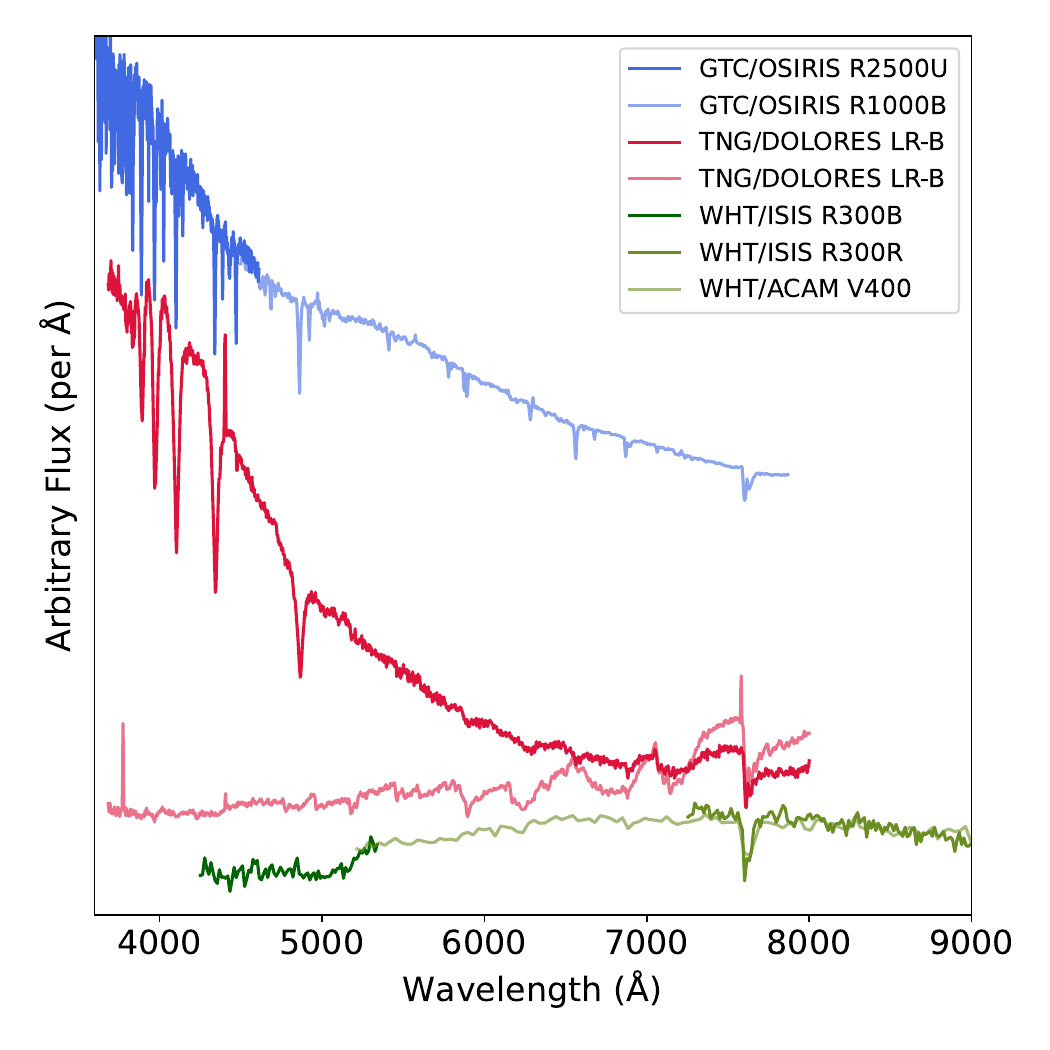}
    \caption{Example extractions of some special cases. Each set of spectra
    uses the same normalization. All spectra are optimally extracted using the
    \citet{1986PASP...98..609H} algorithm. The GTC/OSIRIS (top set in blue hue)
    spectra were flux calibrated in the shorter wavelengths where there are a
    lot of absorption features. With a bad sensitivity curve, we expect many
    small bump features in the spectra and that their continuum would not agree
    in the overlapping wavelength range, as polynomials tend to run away at the
    edges. The TNG/DOLORES (middle set in red hue) spectra were of two targets where
    the tracing was done simultaneously and then the extraction done
    sequentially. The WHT/ISIS and ACAM (bottom set in green hue) data demonstrate
    extracting a target with a very low S/N ratio. The independently
    flux-calibrated spectra agree well with each other in the overlapping
    wavelength ranges.}
    \label{fig:use_cases}
\end{figure}

%%%%%%%%%%%%%%%%%%%%%%%%%%%%%%%%%%%%%%%%%%%%%%%%%%%%%%%%%%%%%%%%%%%%%%%%%%%%%%%%
Aside from the comparison against the independently and externally calibrated
spectra above, we also illustrate a few cases of extractions concerning
other use cases that \textsc{ASPIRED} is capable of dealing with. In the top set
of spectra in Figure~\ref{fig:use_cases}, the GTC/OSIRIS data show a
blue large-amplitude pulsator with many resolved absorption lines in both the
science and the standard targets~(Feige 110\footnote{
\url{https://www.eso.org/sci/observing/tools/standards/spectra/feige110.html}};
~\citealp{2022MNRAS.511.4971M}). With a poorly computed sensitivity curve, we
expect many small bump features in the spectra and that their continuum would
not agree in the overlapping wavelength range, as polynomials tend to run away
at the edges of the fitted range of the wavelength solution. In the middle,
the TNG/DOLORES case shows a simultaneous tracing of a common proper
motion system (three bodies resolved as two components) containing an M dwarf
and an eclipsing M dwarf--white dwarf binary~\citep{2022MNRAS.509.4171K},
separated by 5'' incident on the same long-slit in a single exposure.
At the bottom is a set of three spectra taken with WHT/ISIS and ACAM.
It demonstrates the reduction of a set of data with extremely low S/N
for an ultracool white dwarf with a smooth blackbody-like
spectrum~\citep{2020MNRAS.493.6001L}. The tracing and extraction are only
possible after careful cosmic-ray removal and trimming of the images. The
spectral shapes of the ISIS spectra in both the blue and the red arms are in good
agreement with the ACAM spectrum, and these independently flux-calibrated
spectra agree well with each other in the overlapping wavelength ranges.

%%%%%%%%%%%%%%%%%%%%%%%%%%%%%%%%%%%%%%%%%%%%%%%%%%%%%%%%%%%%%%%%%%%%%%%%%%%%%%%%
To summarize, our test cases cover the most commonly encountered possible
problematic data -- red- or blue-dominated spectra, narrow/broad
absorption/emission lines, featureless spectra, closely spaced (resolved)
line spectra and curved spectra -- and demonstrate that \textsc{ASPIRED} can
produce spectral products at a quality comparable to some
selected published spectra.

%%%%%%%%%%%%%%%%%%%%%%%%%%%%%%%%%%%%%%%%%%%%%%%%%%%%%%%%%%%%%%%%%%%%%%%%%%%%%%%%
\begin{table*}
    \begin{tabular}{l|c|c|c}\hline
        Telescope/Instrument & Object Type                                 & Grating             & Arc \\\hline\hline
        \multicolumn{4}{c}{Conventional Long-slit Setup}\\\hline

        GTC/OSIRIS           & BLAP (ZGP-BLAP-09)                          & R2500U and R1000B    & HgArNe \\
        LT/SPRAT      & Nova shell remnant (DO Aql)                 & 600\,l/mm           & Xe \\
        \multirow{2}{*}{TNG/DOLORES}          & Common proper-motion and eclipsing system     & \multirow{2}{*}{LR-B}                & \multirow{2}{*}{ArKrNeHg} \\
                             & (ZTF J192014.13+272218.1)     &           &  \\
        VLT/FORS             & AM CVn (V418 Ser)                           & 600B                & HeHgCd \\
        WHT/ACAM             & Ultracool white dwarf (PSO J180153.69+625420.24)       & V400                & CuAr \\
        WHT/ISIS             & Ultracool white dwarf (PSO J180153.69+625420.24)       & R300B and R300R      & CuAr \\\hline
        \multicolumn{4}{c}{Other Setup}\\\hline
        Las Cumbres/FLOYDS           & Type II-P (iPTF14hls)                       & 235\,l/mm           & HgArZn \\
        Gemini/GMOS          & Kilonova (AT 2017gfo)                       & R400 and B600        & CuAr \\\hline
\end{tabular}
    \caption{List of the eight targets used in Figures~\ref{fig:reduction_compared} \& \ref{fig:use_cases} in alphabetical order of the telescope names. See body text for references.}
    \label{tab:summary}
\end{table*}

%%%%%%%%%%%%%%%%%%%%%%%%%%%%%%%%%%%%%%%%%%%%%%%%%%%%%%%%%%%%%%%%%%%%%%%%%%%%%%%%

\subsection{Repeatability}
\begin{figure}
    \centering
    \includegraphics[width=\columnwidth]{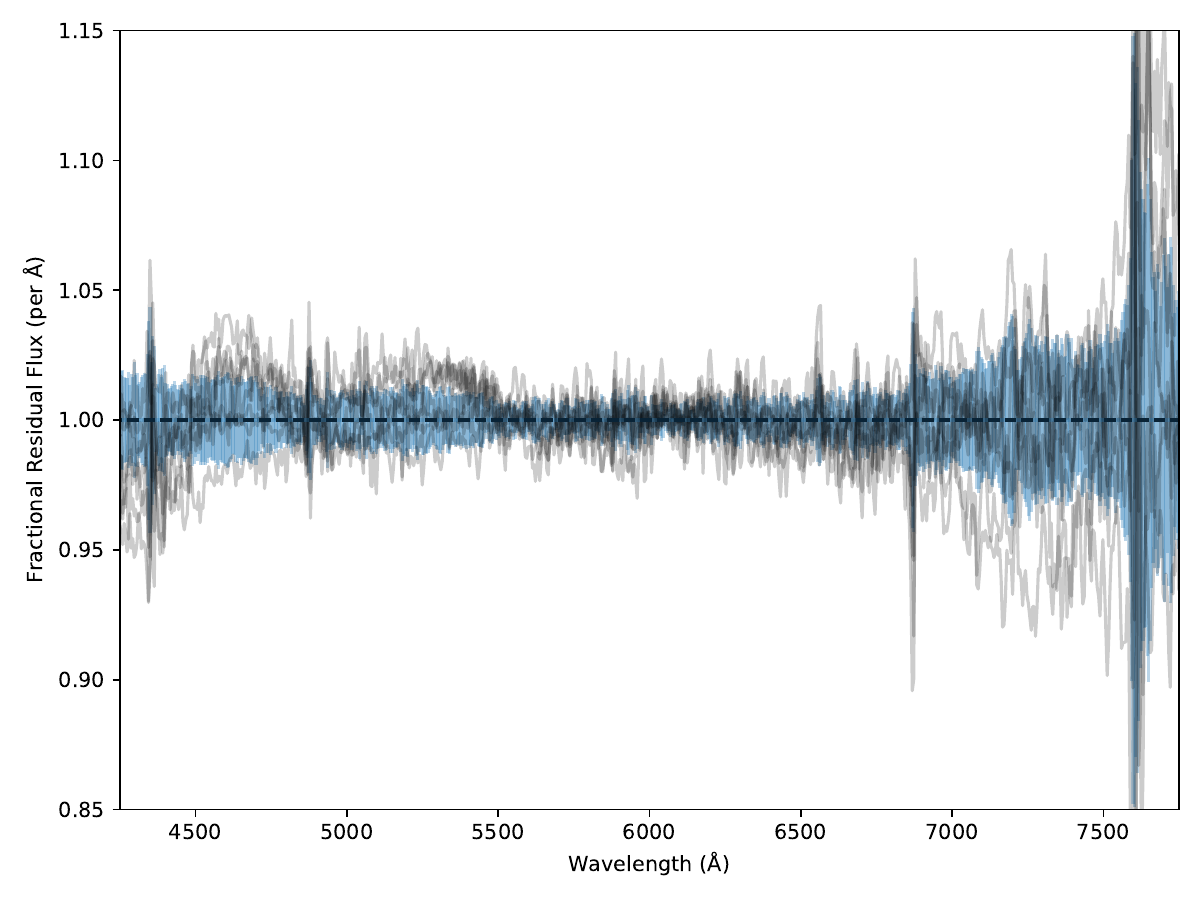}
    \caption{Fractional residual in the calibrated flux of BD\,33+2642 using
    Hiltner\,102 as a standard star taken with LT/SPRAT. Each gray
    line depicts the calibration residual between the median spectrum of the eight spectra
    taken on different nights. The blue area shows the one standard deviation
    of the scatter in the residual. The spikes above 5\% residual are at the
    H$\gamma$ line at 4340\,\AA\ and the oxygen telluric features at 6869 and
    7605\,\AA\ . The residuals that are larger on the two ends are expected as
    the S/N ratios drop with the detector efficiency on either
    side.}
    \label{fig:repeatability}
\end{figure}

In order to demonstrate reproducibility and repeatability,
we opt to calibrate the standard star BD33+2642 against Hiltner\,102 on
eight nights for which LT/SPRAT observed both on the same night during
semester 2022\,B. We require the transparency to be photometric, and the seeing
be better than 2'' as reported by the autoguider. Because of the robotic
observing nature of the LT, standard stars are only taken at the beginning
and at the end of the night. The data used were not taken immediately
before or after each other, and therefore weather effects cannot be calibrated out.
We only apply atmospheric extinction corrections and telluric
absorption corrections. The spectra are consistent with each other to within
1\%-2\%~(apart from the telluric absorption regions and at the H$\gamma$ line at 4340\,\AA\
that are observed at low detector efficiency; see Fig.~\ref{fig:repeatability}).

%%%%%%%%%%%%%%%%%%%%%%%%%%%%%%%%%%%%%%%%%%%%%%%%%%%%%%%%%%%%%%%%%%%%%%%%%%%%%%%%
\section{Distribution}
\label{sec:distribution}

\textsc{ASPIRED} is released under the BSD~(3-Clause) License. The source code
is hosted on \textsc{Github}. The DOI of each version can be found on 
\textsc{zenodo}: \url{https://zenodo.org/record/4127294}. For more
straightforward installation, \textsc{ASPIRED} is also available with the
\textsc{Python Package Index}~(\textsc{PyPI}) so users can install the software with a single command: 
\begin{verbatim}
>> pip install aspired
\end{verbatim}
The latest stable version can be installed with
\begin{verbatim}
>> pip install git+https://github.com/
    cylammarco/ASPIRED@main
\end{verbatim}
and the development version can be installed with 
\begin{verbatim}
>> pip install git+https://github.com/
    cylammarco/ASPIRED@dev-v[Major].[Minor].X
\end{verbatim} Here, \verb+[Major]+ and \verb+[Minor]+ are the major and minor
version numbers. \verb+X+ is an actual character of the branch name, as we only
use separate maintenance branches at the level of minor version granularity.

It is also possible to clone the entire repository
and install with the setup script using
\begin{verbatim}
>> git clone [url to main/dev/a specific commit]
>> python setup.py
\end{verbatim}
.

All the reduction scripts for the spectra used in this article can be found online
\footnote{\url{https://github.com/cylammarco/ASPIRED-apj-article}}. There are also example
reduction scripts available from the companion repository
\textsc{ASPIRED-example}\footnote{\url{https://github.com/cylammarco/ASPIRED-example}}
which are tagged by version number starting from \texttt{v0.4}. This work is most
accurately reflects the software release of the \texttt{v0.5}-series.
We encourage users to reduce example spectra available in these repositories, using their own pipelines, for comparison.

%%%%%%%%%%%%%%%%%%%%%%%%%%%%%%%%%%%%%%%%%%%%%%%%%%%%%%%%%%%%%%%%%%%%%%%%%%%%%%%%
\section{Summary and Future work}
\label{sec:summary}

We expect to continue the development and maintenance of \textsc{ASPIRED} in the
future. Some of the main planned developments include the following:
\begin{enumerate}
    \item Improving the quality of the flux calibration by
    carefully masking out the absorption lines in the standard spectra, and
    treating the high- and low-resolution templates differently.
    \item Accepting arcs taken at different times and using the effective wavelength
    calibration solution over a long observing block.
    \item Accepting standards taken at different times and computing the effective
    sensitivity function that takes into account the nonnegligible change
    in the flux of the target over a long observing block.
    \item Exporting reduced spectra as data type \texttt{specutil.spectrum1D}
    for smooth integration to the \textsc{Astropy} analytical tools.
    \item Improving static image output support.
    \item Supporting more forms of optimal extractions.
\end{enumerate}

We anticipate that ASPIRED will be useful for many individual researchers as
well as observatories. It offers a complete set of methods in \textsc{Python}
for spectral data reduction from raw images to final products. It sits between
\textsc{specreduce}, which provides low-level reduction functions, and 
\textsc{PypeIt}, which offers preconfigured tailor-made reduction methods for
specific instrument configuration. \textsc{ASPIRED}'s simple structure and
syntax, with a rich set of examples and documentation, is aimed to make the
learning curve as gentle as possible. The relatively lightweight and
cross-platform design allow a wide audience, including those with less
computing equipment and power, to perform all necessary reduction tasks.

%%%%%%%%%%%%%%%%%%%%%%%%%%%%%%%%%%%%%%%%%%%%%%%%%%%%%%%%%%%%%%%%%%%%%%%%%%%%%%%%
\section*{Acknowledgements}
This work was partially supported by OPTICON. This project has
received funding from the European Union's Horizon 2020 research and
innovation program under grant agreement No 730890. This material
reflects only the authors' views and the Commission is not liable for
any use that may be made of the information contained therein.

This work was partially supported by the Polish NCN grant Daina
No. 2017/27/L/ST9/03221.

M.C.L. is supported by a European Research Council (ERC) grant under the European Union's Horizon 2020 research and innovation program (grant agreement number 852097).

I.A. is a CIFAR Azrieli Global Scholar in the Gravity and the Extreme Universe Program and acknowledges support from that program, from the ERC under the European Union's Horizon 2020 research and innovation program (grant agreement number 852097), from the Israel Science Foundation (grant No. 2752/19), from the United States -- Israel Binational Science Foundation (BSF), and from the Israeli Council for Higher Education Alon Fellowship.

The LT is operated on the island of La Palma by Liverpool
John Moores University in the Spanish Observatorio del Roque
de los Muchachos of the Instituto de Astrof{\'i}sica de Canarias with
financial support from the UK Science and Technology Facilities
Council.

This work makes use of observations from the Las Cumbres Observatory
global telescope network. We have made use of the data collected from
the FLOYDS spectrograph on the LCOGT 2m telescope at both Siding Spring,
Australia and Maui, HI, United States.

The William Herschel Telescope and its service program are operated
on the island of La Palma by the Isaac Newton Group of Telescopes in
the Spanish Observatorio del Roque de los Muchachos of the Instituto
de Astrof\'isica de Canarias.

Based on observations made with the Gran Telescopio Canarias~(GTC), installed 
in the Spanish Observatorio del Roque de los Muchachos of the Instituto de 
Astrof\'isica de Canarias, in the island of La Palma.

Based on observations obtained at the Gemini Observatory, which is operated by the
Association of Universities for Research in Astronomy, Inc., under a cooperative agreement
with the NSF on behalf of the Gemini partnership: the National Science Foundation (United
States), the Science and Technology Facilities Council (United Kingdom), the
National Research Council (Canada), CONICYT (Chile), the Australian Research Council
(Australia), Ministério da Ciência e Tecnologia (Brazil), and SECYT (Argentina).

Based on observations collected at the European Southern Observatory under ESO program(s) 095.D-0888(D).

%%%%%%%%%%%%%%%%% APPENDICES %%%%%%%%%%%%%%%%%%%%%

\software{
    \textsc{astro-scrappy}~\citep{2001PASP..113.1420V, curtis_mccully_2018_1482019}
    \textsc{Astropy}~\citep{astropy:2013, astropy:2018},
    \textsc{ccdproc}~\citep{matt_craig_2017_1069648},
    \textsc{matplotlib}~\citep{Hunter:2007},
    \textsc{numpy}~\citep{2020NumPy-Array}
    \textsc{plotly}~\citep{plotly},
    \textsc{rascal}~\citep{2020ASPC..527..627V},
    \textsc{scipy}~\citep{2020SciPy-NMeth},
    \textsc{spectresc}~\citep{2017arXiv170505165C, lam_marco_c_2023_7865549},
    \textsc{specutil}~\citep{nicholas_earl_2023_7803739}, and
    \textsc{statsmodels}~\citep{seabold2010statsmodels}
    }

\appendix
\section{Code Snippets}
\label{sec:appendix}
The following compares the reduction scripts for LT/SPRAT and
Gemini/GMOS reduction. LT/SPRAT starts from an \texttt{ImageReduction} object,
while the Gemini/GMOS workflow uses flattened images in FITS objects.
Both have all the necessary header information for the reduction.
The examples are simplified to exclude any file- and image-saving
instructions. They are shown in three groups according to the types
of operation:

(i)~tracing and extracting the spectra from two-dimensional images:

\vspace*{2em}
\noindent
\begin{minipage}{0.475\linewidth}
\begin{Verbatim}[frame=topline,numbers=left,label=LT/SPRAT,framesep=3mm]
# Initialise the two TwoDSpec() for the
# science and standard. The arc frames
# in the objects are handled automatically
science_twodspec = TwoDSpec(
    science_image_reduction_object,
    spatial_mask=sprat_spatial_mask,
    spec_mask=sprat_spec_mask,
)
standard_twodspec = TwoDSpec(
    standard_image_reduction_object,
    spatial_mask=sprat_spatial_mask,
    spec_mask=sprat_spec_mask,
)

# Automatically trace the spectrum
science_twodspec.ap_trace()
standard_twodspec.ap_trace()

# Tophat extraction of the spectrum
science_twodspec.ap_extract(
    apwidth=12,
    skysep=12,
    skywidth=5,
    skydeg=1,
    optimal=False,
)
# Optimal extracting spectrum
standard_twodspec.ap_extract(
    apwidth=15,
    skysep=3,
    skywidth=5,
    skydeg=1,
)

# Extract the 1D arc by aperture sum of the
# traces provided
science_twodspec.extract_arc_spec()
standard_twodspec.extract_arc_spec()
\end{Verbatim}
\end{minipage}\hfill
\begin{minipage}{0.475\linewidth}
\begin{Verbatim}[frame=topline,numbers=left,label=Gemini/GMOS,framesep=3mm]
# Initialise the two TwoDSpec() for the
# science and standard
science_twodspec = TwoDSpec(
    science_fits_object,
    spatial_mask=gmos_spatial_mask,
    spec_mask=gmos_spec_mask,
    flip=True,
)
standard_twodspec = TwoDSpec(
    standard_fits_object,
    spatial_mask=gmos_spatial_mask,
    spec_mask=gmos_spec_mask,
    flip=True,
)

# Automatically trace the spectrum
science_twodspec.ap_trace(fit_deg=2)
standard_twodspec.ap_trace(fit_deg=2)

# Extracting of the spectrum
science_twodspec.ap_extract()
standard_twodspec.ap_extract(
    apwidth=25,
)

# Handling the arc frame
science_twodspec.add_arc(science_arc)

# Use apply_mask to handle the flip
science_twodspec.apply_mask_to_arc()
science_twodspec.extract_arc_spec()

# Alternatively, it can be flipped manually
standard_twodspec.add_arc(
    np.flip(standard_arc)
)
standard_twodspec.extract_arc_spec()

\end{Verbatim}
\end{minipage}

\clearpage

(ii)~wavelength calibration, where the \texttt{effective\_peaks}
and \texttt{science\_pixel\_list} in the GMOS reduction is
to account for the pixel positions due to chip gaps;

\vspace*{1em}

\noindent
\begin{minipage}{0.475\linewidth}
\begin{Verbatim}[frame=topline,numbers=left,label=LT/SPRAT,framesep=3mm]
# Initialise an OneDSpec() for the
# science and standard targets
spec = OneDSpec()
spec.from_twodspec(
    science_twodspec,
    stype="science"
)
spec.from_twodspec(
    standard_twodspec,
    stype="standard"
)

# Find the peaks of the arc
spec.find_arc_lines(
    top_n_peaks=35,
    prominence=1,
)

# Configure the wavelength calibrator
spec.initialise_calibrator()
spec.set_hough_properties(
    num_slopes=500,
    xbins=100,
    ybins=100,
    min_wavelength=3500,
    max_wavelength=8000,
    range_tolerance=250,
)

# Add user-provided atlas
spec.add_user_atlas(
    elements=sprat_element,
    wavelengths=sprat_atlas,
)

# More configuration of the calibrator
spec.do_hough_transform()
spec.set_ransac_properties(
    minimum_matches=18,
)

# Fit for the wavelength solution
spec.fit(max_tries=1000)

# Apply the fitted solution
spec.apply_wavelength_calibration()




\end{Verbatim}
\end{minipage}\hfill
\begin{minipage}{0.475\linewidth}
\begin{Verbatim}[frame=topline,numbers=left,label=Gemini/GMOS,framesep=3mm]
# Initialise an OneDSpec() for the
# science and standard targets
spec = OneDSpec()
spec.from_twodspec(
    science_twodspec,
    stype="science",
)
spec.from_twodspec(
    standard_twodspec,
    stype="standard",
)

# Find the peaks of the arc
spec.find_arc_lines(prominence=0.25)

# Configure the wavelength calibrator
spec.initialise_calibrator(
    peaks=effective_peaks,
)
spec.set_calibrator_properties(
    pixel_list=science_pixel_list,
)
spec.set_hough_properties(
    num_slopes=10000.0,
    xbins=500,
    ybins=500,
    min_wavelength=4900.0,
    max_wavelength=9500.0,
)
spec.add_user_atlas(
    elements=gmos_elements,
    wavelengths=gmos_atlas_lines,
    pressure=gmos_pressure,
    temperature=gmos_temperature,
    relative_humidity=gmos_rh,
)
spec.do_hough_transform()
spec.set_ransac_properties(
    sample_size=6,
    minimum_matches=15, 
)

# Fit for the wavelength solution
spec.fit(
    max_tries=1000,
    fit_deg=4,
    fit_tolerance=10.0,
)
# Apply the fitted solution
spec.apply_wavelength_calibration()
\end{Verbatim}
\end{minipage}

and (iii)~flux calibration, atmospheric extinction correction and telluric absorption correction.

\vspace*{1em}

\noindent
\begin{minipage}{0.475\linewidth}
\begin{Verbatim}[frame=topline,numbers=left,label=LT/SPRAT,framesep=3mm]
# Choose the standard star
spec.load_standard(
    target="BDp332642",
    library="irafirscal",
)

# Compute the sensitivity/response function
spec.get_sensitivity()

# Accept the sensitivity/response function
spec.apply_flux_calibration()

# Correct for atmospheric extinction
spec.set_atmospheric_extinction(
    location="orm"
)
spec.apply_atmospheric_extinction_correction()

# Telluric absorption correction
spec.get_telluric_profile()
spec.get_telluric_strength()
spec.apply_telluric_correction()

# Resample the spectra
spec.resample()
\end{Verbatim}
\end{minipage}\hfill
\begin{minipage}{0.475\linewidth}
\begin{Verbatim}[frame=topline,numbers=left,label=Gemini/GMOS,framesep=3mm]
# Choose the standard star
spec.load_standard(
    target="EG274",
    library="esoxshooter",
)

# Compute the sensitivity/response function
spec.get_sensitivity()

# Accept the sensitivity/response function
spec.apply_flux_calibration(stype="science")

# Correct for atmospheric extinction
spec.set_atmospheric_extinction(location="ls")
spec.apply_atmospheric_extinction_correction()

# Telluric absorption correction
spec.get_telluric_profile()
spec.get_telluric_strength()
spec.apply_telluric_correction()

# Resample the spectra
spec.resample()


\end{Verbatim}
\end{minipage}


\bibliography{main}{}
\bibliographystyle{aasjournal}



\end{document}